\documentclass[oldversion]{aa}
\usepackage{graphicx}
\usepackage{natbib}
\usepackage{url}
\usepackage{ulem}
\usepackage[T1]{fontenc}
\usepackage{ae,aecompl}
\usepackage{times}
\usepackage{pslatex}
\usepackage{bm}
\usepackage{upgreek}
\makeatletter
\newcommand{\bfgreek}[1]{\bm{\@nameuse{up#1}}}
\newcommand{\micron }{\mbox{$\bfgreek{mu}$m }}
\newcommand{\microK }{\mbox{$\bfgreek{mu}$K }}
\newcommand{\microV }{\mbox{$\bfgreek{mu}$V}}
      \def\new#1 {{\bf #1 }}  
      \def\cut#1 {\sout{#1} }

\bibpunct{(}{)}{;}{a}{}{,}
\usepackage{txfonts}
\begin{document}
\title{The Large APEX Bolometer Camera LABOCA}

\author{
   G. Siringo\inst{1},
   E. Kreysa\inst{1},
   \\
   A. Kov{\' a}cs\inst{1},
   F. Schuller\inst{1},
   A. Wei\ss\inst{1},
   \\
   W. Esch\inst{1},
   H.-P. Gem\"und\inst{1},
   N. Jethava\inst{2},
   G. Lundershausen\inst{1},
   A. Colin\inst{3},
   \\
   R. G\"usten\inst{1},
   K. M. Menten\inst{1},
   A. Beelen\inst{4},
   F. Bertoldi\inst{5},
   J. W. Beeman\inst{6},
   E. E. Haller\inst{6},
   }

\offprints{G. Siringo \\ \email{gsiringo@mpifr-bonn.mpg.de}}

\institute{Max-Planck-Institut f\"ur Radioastronomie, Auf dem H\"ugel 69, 53121 Bonn, Germany
\and
National Institute of Standards and Technology, Boulder, CO 80305, USA
\and
Instituto de Fisica de Cantabria (CSIC-UC), Avda. Los Castros, 39005 Santander, Spain
\and
Institut d'Astrophysique Spatiale, b\^at 121 - Universit\'e Paris-Sud, 91405 Orsay Cedex, France
\and
Argelander-Institut f\"ur Astronomie, University of Bonn, Auf dem H\"ugel 71, 53121 Bonn, Germany
\and
Lawrence Berkeley National Laboratory, Berkeley, CA 94720, USA
}

\date{Received 2 December 2008 / Accepted 21 January 2009}

\abstract
{
The Large APEX Bolometer Camera, LABOCA, has been commissioned for operation as a new facility instrument
at the Atacama Pathfinder Experiment 12\,m submillimeter telescope.
This new 295-bolometer total power camera, operating in the 870\,\micron atmospheric window,
combined with the high efficiency of APEX and the excellent atmospheric transmission at the site,
offers unprecedented capability in mapping submillimeter continuum emission for a wide range of astronomical purposes.
}

\keywords{Instrumentation: detectors -- Instrumentation: photometers -- Techniques: bolometers -- Techniques: array -- Techniques: submillimeter}
\authorrunning{G. Siringo et al.}
\titlerunning{The Large APEX Bolometer Camera LABOCA}
\maketitle
%
%
\section{Introduction}
\subsection{Astronomical motivation}\label{sec:intro} 
Millimeter and submillimeter wavelength continuum emission 
is a powerful probe of the warm and cool dust in the Universe.
For temperatures below $\sim$40\,K, the peak of the thermal continuum emission is at wavelengths longer than 100\,\micron 
(or at frequencies lower than 3\,THz), i.e.~in the far-infrared and (sub)millimeter\footnote{Hereafter we will use the term {\it (sub)millimeter} when referring, in general, to the millimeter and submillimeter wavelengths range;
we will use {\it submillimeter} when strictly referring to wavelengths from one millimeter down to the far-infrared.} range.
A number of {\it atmospheric windows} between 200\,GHz and 1\,THz make ground-based observations possible over a large part of
this range from high-altitude, dry sites.

Specifically, thermal dust emission is well described by a gray body spectrum,
with the measured flux density $S_\nu$ at frequency $\nu$ expressed as:
\[ 
S_\nu = \Omega_{S'} ~ (1-e^{-\tau}) ~ \frac{2h}{c^2}
\frac{ \nu^3 } {e^{h\nu / kT} - 1}
\]
where $h$ and $k$ are Planck's and Boltzmann's constants respectively,
$c$ is the speed of light, $\Omega_{S'}$ is the apparent source solid angle
(the size of the physical source convolved with the telescope beam)
and $\tau$ is the optical depth, which varies with frequency.
In the (sub)millimeter range, the emission is almost always optically thin with $\tau \propto N \nu^\beta$,
where $\beta$ is in the range 1--2 typically
(see, e.g.,~the Appendix of \citealt{Mezger1990} or \citealt{Beuther2002} for a more thorough discussion).
Here, $N$ is the number of dust particles (or number of nucleons assuming a given dust-to-gas
relation) in the telescope beam. Accordingly, the (sub)millimeter flux
can be converted into dust/gas masses, when the temperature $T$ is
assumed or constrained via additional far-infrared measurements.

Such mass determination is one of the core issues of (sub)millimeter
photometry. We would like to illustrate its paramount importance with
a few examples: (sub)millimeter wavelengths mapping of low-mass
star-forming regions in molecular clouds have determined the dense
core mass spectrum, (in nearby regions) down to sub-stellar masses,
and investigated its relationship to the Initial Mass Function
\citep{Motte1998}.  These studies can be extended to high-mass
star-forming regions \citep[e.g.~][]{Motte2007,
Johnstone2006}. Because of the larger distances to rarer, high-mass
embedded objects, even relatively shallow surveys are capable of
detecting pre-stellar cluster clumps with a few hundreds of solar
masses of material at distances as far as the Galactic center. Masses
and observed sizes yield radial density distribution profiles for
protostellar cores that can be compared with theoretical models
\citep{Beuther2002}.

(Sub)millimeter continuum emission is also a remarkable tool for the
study of the distant Universe. The thermal dust emission in active 
galaxies is typically fueled by short-lived, high-mass stars, therefore
the far-infrared luminosity provides a snapshot of the current level of star-formation activity.
Deep (sub)millimeter observations in
the Hubble Deep Field, with the Submillimeter Common User Bolometer
Array \citep[SCUBA,][]{1999MNRAS.303..659H} on the James Clerk Maxwell
Telescope, attracted considerable attention with the detection of a
few sources without optical or near-infrared counterpart.  Sensitive
measurements found dust in high redshift sources and even in some of
the farthest known objects in the Universe \citep[see, e.g.,][]{Carilli2001, Bertoldi2003, Wang2008},
revealing star-formation rates (SFRs) that are hundreds of times higher than in
the Milky Way today. These detections are possible because of the so-called
{\it negative K-correction} first discussed by \citet{Blain1993}: the
warm dust in galaxies is typically characterized by temperatures
around 30--60\,K. Its thermal emission dominates the spectral energy
distribution (SED) of luminous and ultra-luminous infrared galaxies
(LIRGs and ULIRGs), which have maxima between 3 and 6\,THz as a
result. Because of the expansion of the Universe, the peak of the
emission shifts toward the lower frequencies with increasing
cosmological distance, thus counteracting the dimming and benefitting 
detection at (sub)millimeter wavelength. Consequently,
flux-limited surveys near millimeter wavelengths yield flat or nearly
flat luminosity selection over much of the volume of the Universe
\citep{Blain02}. As such, (sub)millimeter wavelengths allow unbiased
studies of the luminosity evolution and, therefore, of the star-formation
history of galaxies over cosmological time-scales.
\subsection{Bolometer arrays for (sub)millimeter astronomy}
\begin{figure}[t]
\centering 
\includegraphics[width=.96\linewidth, bb=20 20 1364 1316, clip]{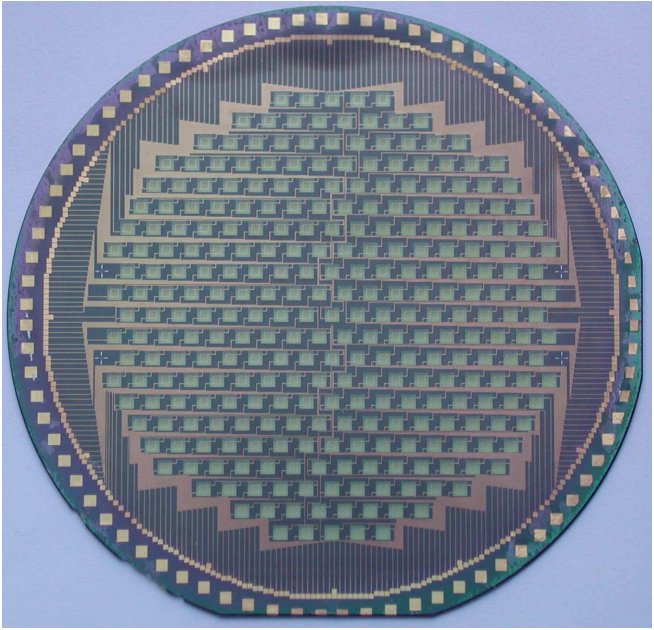}
\caption
{
Wiring side of a naked LABOCA array. Each light-green square is a bolometer.
}
\label{fig:nakedarray}
\end{figure}
The (sub)millimeter dust emission is unfortunately intrinsically weak and
the wish to measure it pushed the development of detectors with the best
possible sensitivity, namely bolometers \citep{Low1961, Mather1984}.
Moreover, the desirability of mapping large areas of the sky,
motivated the development of detector {\it arrays}.
Consequently, the last 10 years have seen an
increasing effort in the development of bolometer arrays.  In such
instruments, a number of composite bolometers work side by side in the
focal plane, offering simultaneous multi-beam coverage.  Since the
arrival of the first arrays, developed in the early '90s and
consisting of just 7 elements, we are witnessing a rapid maturing of
technology, reaching hundreds to a few thousand elements today, and
the prospect of even larger bolometer arrays in the future.
The Large APEX Bolometer Camera, LABOCA, described in this article,
is a new bolometric receiver array for the Atacama Pathfinder Experiment 12\,m telescope,
APEX\footnote{APEX is a collaborative effort between the
Max-Planck-Institut f\"ur Radioastronomie of Bonn (MPIfR,\,50\%), the
European Southern Observatory (ESO,\,27\%) and the Onsala Space
Observatory (OSO,\,23\%)}.  It is the most ambitious camera in a long
line of developments of the MPIfR bolometer group, which has delivered
instruments of increasing complexity to the IRAM 30 m telescope
\citep{Kreysa1990}: single beam receivers were supplanted by a
7-element system \citep{Kreysa1999}, which eventually gave way to the
MAx-Planck Millimeter BOlometer (MAMBO) array, whose initial 37 beams
have grown to 117 in the latest incarnation \citep{2002AIPC..616..262K}.
The group also built the 37-element 1.2\,mm SEST Imaging Bolometer Array (SIMBA)
for the Swedish/ESO Submillimeter telescope \citep{Nyman2001}
and the 19-element 870\,\micron  for the 10\,m Heinrich-Hertz Telescope
\citep[a.k.a. SMTO,][]{1990SPIE.1235..503M}.

The main obstacle to observations at these wavelengths is posed by Earth's atmosphere, which is
seen as a bright emitting screen by a continuum total power detector, as LABOCA's bolometers are.
This is largely due to the emission of the water vapor present in the atmosphere with only small
contributions from other components, like ozone. Besides, the atmosphere is a turbulent thermodynamic
system and the amount of water vapor along the line of sight can change quickly, giving rise to instabilities
of emission and transmission, called {\it sky noise}. Observations from ground based telescopes have to go
through that screen, therefore requiring techniques to minimize those effects.

The technique most widely used is to operate a switching device,
usually a chopping secondary mirror (hereafter called {\it wobbler})
to observe alternatively the source and a blank sky area close to it,
at a frequency higher than the variability scale of the sky noise.
This method, originally introduced for observations with single pixel detectors, is today also used with arrays of bolometers.
Although it can be very efficient to reduce the atmospheric disturbances during observations,
it presents some disadvantages: among others, the wobbler is usually slow (1 or 2 Hz) posing a
strong limitation to the possible scanning speed.

LABOCA has been specifically designed to work without a wobbler and using a different technique,
which works particularly well when using an array of detectors, to remove the atmospheric contribution.
This technique, called {\it fast scanning} \citep{2001A&A...379..735R}, is based on the idea that, when observing
with an array, the bolometers composing the array look simultaneously at different points in the sky,
therefore chopping is no more needed.
A modulation of the signal, still required to identify the astronomical source through the atmospheric
emission, is produced by scanning with the telescope across the source.
The atmospheric contribution (as well as part of the instrumental noise) will be strongly correlated in
all bolometers and a post-detection analysis of the correlation across the array will 
make it possible to extract the signals of astronomical interest from the atmospheric foregrounds.
Moreover, the post-detection bandwidth depends on the scanning speed, therefore relatively high
scanning speeds are ideal \citep[see also][]{Kovacs2008}.

The APEX telescope \citep{2006A&A...454L..13G}, as the name implies,
serves as a pathfinder for the future large-scale (sub)millimeter
wavelength and (far)infrared missions, namely the Atacama Large
Millimeter Array (ALMA), the Herschel Space Observatory and the
Stratospheric Observatory for Infrared Astronomy (SOFIA). Its
pathfinder character is on the one hand defined by exploring
wavelength windows that have been poorly studied before, with
acceptable atmospheric transmission at the 5100\,m altitude site.  On
the other hand, and more importantly, it can perform large area
mapping to identify interesting sources for ALMA follow-up studies at
higher angular resolution.  Moreover, APEX produces images of both
continuum (with LABOCA) and line emission with angular resolutions
that neither Herschel's nor SOFIA's smaller telescopes (with diameters
of 3.5 and 2.5\,m respectively) can match.  This provides a critical
advantage to APEX for imaging dust and line emission at high frequencies. 
%
%
\begin{figure*}[ht]
\centering
\includegraphics[width=.90\linewidth,bb=22 22 699 535, clip]{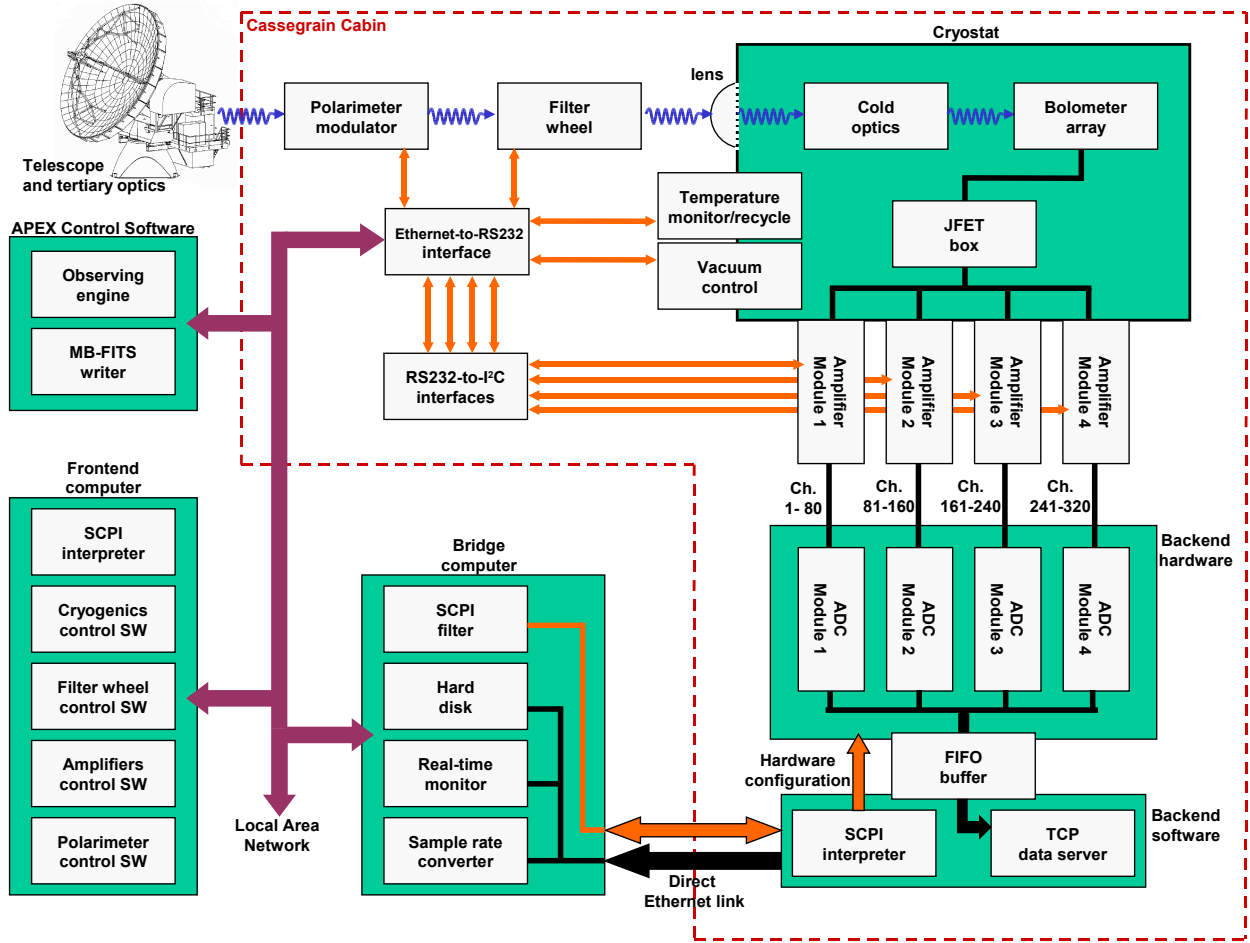}
\caption
{
Scheme of the infrastructure of LABOCA. The instrument is located in the Cassegrain cabin (dashed red line)
of APEX, remote operation is a requirement and all the communication goes through
the local area network (magenta arrows) and a direct Ethernet link between {\it backend} and {\it bridge} computers.
The black lines show the flow of the bolometer signals, the orange ones show the configuration and monitoring communication.
}
\label{fig:infrastructure}
\end{figure*}
\subsection{Instrument overview}\label{sec:overview}
LABOCA is an array of bolometers, operated in total-power mode, specifically designed 
for fast mapping of large areas of sky at moderate resolution and with high 
sensitivity and was commissioned in May 2007 as facility instrument on APEX.
It is a very complex system, composed of parts that originate in a variety of
fields of technology, in particular optics, high vacuum,
low temperature cryogenics, digital electronics, computer hardware and
software, and others.  A general view of the infrastructure is shown
in the block diagram of Fig.~\ref{fig:infrastructure}.
The heart of LABOCA is its detector array made of 295 semiconducting composite bolometers
(see Fig.~\ref{fig:nakedarray}, Fig.~\ref{fig:arraypics}).
A description of the detector array design and manufacture is provided in \S\ref{sec:detector}.

The bolometer array is mounted in a cryostat, which uses liquid
nitrogen and liquid helium on a closed cycle double-stage sorption
cooler to reach an operation temperature of $\sim$285\,mK. The
cryogenic system is discussed in \S\ref{sec:cryo}.

A set of cold filters, mounted on the liquid nitrogen and liquid
helium shields, define the spectral passband, centered at a wavelength
of 870\,\micron  and about 150\,\micron  wide (see
Fig.~\ref{fig:passband}).  A monolithic array of conical horn
antennas, placed in front of the bolometer wafer, concentrates the
radiation onto the individual bolometers.  The filters and the horn
array are presented in \S\ref{sec:coptics}.

The cryostat is located in the Cassegrain cabin of the APEX telescope
(see Fig.~\ref{fig:receiver}) and the optical coupling to the
telescope is provided by an optical system made of a series of metal
mirrors and a lens which forms the cryostat window.  The complex
optics layout, manufacture and installation at the telescope in
described in \S\ref{sec:optics}.

The bolometer signals are routed through low noise, unity gain
Junction Field Effect Transistor (JFET) amplifiers and to the outside
of the cryostat along flexible flat cables.  Upon exiting the
cryostat, the signals pass to room temperature low noise amplifiers
and electronics also providing the AC current for biasing the
bolometers and performing real time demodulation of the signals.  The
signals are then digitized over 16 bits by 4 data acquisition boards
providing 80 analog inputs each, mounted in the {\it backend
computer}. The backend software provides an interface to the
telescope's control software, used to set up the hardware, and a data
server for the data output.  The acquired data are then digitally
filtered and downsampled to a lower rate in real time by the {\it
bridge computer} and finally stored in MB-FITS format
\citep{2006A&A...454L..25M} by the FITS writer embedded in the
telescope's control software.  Another computer, the {\it frontend
computer}, is devoted to monitor and control most of the
electronics embedded into the receiver (e.g.~monitoring of all the
temperature stages, controlling of the sorption cooler, calibration
unit) and provides an interface to the APEX control software, allowing
remote operation of the system.  Discussions of the cold and warm
readout electronics, plus the signal processing, are found in
\S\ref{sec:jfet} and \S\ref{sec:electronics}, respectively.
\begin{figure*}[ht]
\centering
\includegraphics[width=.92\linewidth, bb= 20 20 1147 846,clip]{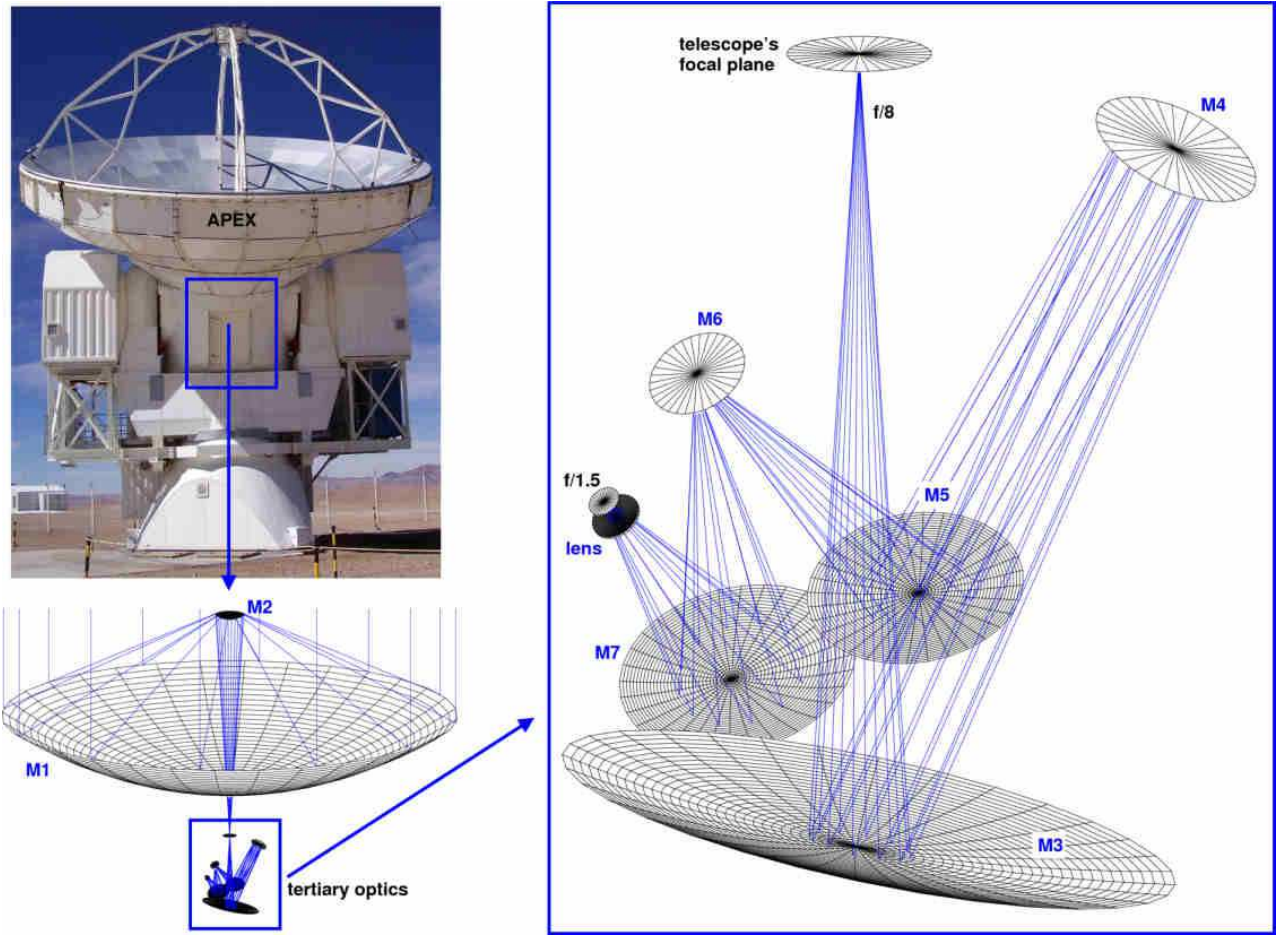}
\caption
{
Overview of the optics. On the left, the APEX telescope.
On the right, a zoom on the tertiary optics installed in the Cassegrain cabin.
See also Fig.~\ref{fig:receiver}.
}
\label{fig:opticsall}
\end{figure*}

In \S\ref{sec:obsmodes} we describe the
sophisticated observing techniques used with LABOCA,
some of them newly developed. 
The instrument performance on the sky is described in \S\ref{sec:performance},
along with information on sensitivity, beam shape and noise behavior.

The reduction of the data is performed using a new data reduction
software included in the delivery of LABOCA as facility instrument,
the BoA (Bolometer array data Analysis) data reduction software package.
An account of on-line and off-line data reduction is given in \S\ref{sec:redu}.

Some of the exciting science results already obtained with LABOCA or
expected in the near future are outlined in \S\ref{sec:science}. Our
plans for LABOCA's future are briefly presented in \S\ref{sec:future}.
%
%
\section{Tertiary optics}\label{sec:optics}
\subsection{Design and optimization} 
The very restricted space in the Cassegrain cabin of
APEX and a common first mirror (M3) with the APEX SZ Camera
\citep[ASZCa,][]{2003NewAR..47..933S} introduced many boundary
conditions into the optical design. Eventually, with the help of the
ZEMAX\footnote{\tiny{\url{http://www.zemax.com/}}} optical design program, a
satisfactory solution was found, featuring three aspherical off-axis
mirrors (M3, M5, M7), two plane mirrors (M4, M6) and an aspherical
lens acting as the entrance window of the cryostat (see
Fig.~\ref{fig:opticsall}). Meeting the spatial constraints, without
sacrificing optical quality, is facilitated considerably by the
addition of plane mirrors.
The design of the optics was made at the MPIfR in coordination with N.\ Halverson\footnote{Formerly
at University of California at Berkeley, now at University of
Colorado, Boulder, USA} with respect to sharing the large M3 mirror
with the ASZCa experiment.

The maximum possible field diameter of APEX, as limited by the
diameter of the Cassegrain hole of the telescope's primary, is about
0.5 degrees. LABOCA, with its 295 close-packed fully efficient horns,
covers an almost circular field of view (hereafter FoV) of about 0.2
degrees in diameter (or about 100 square arcminutes).  The task of the
tertiary optics is to transform the f-ratio from f/8 at the Cassegrain
focus to f/1.5 at the horn array, while correcting the aberrations
over the whole FoV of LABOCA under the constraint of parallel output
beams.  The final design is diffraction limited even for 350\,\micron 
wavelength, the Strehl ratio is better than 0.994 and the maximum
distortion at the focal plane is less than 10\% over the entire FoV
(see also Fig.~\ref{fig:beams}).
\begin{figure}[t] 
\centering 
\includegraphics[width=.95\linewidth,bb=20 20 575 436, clip]{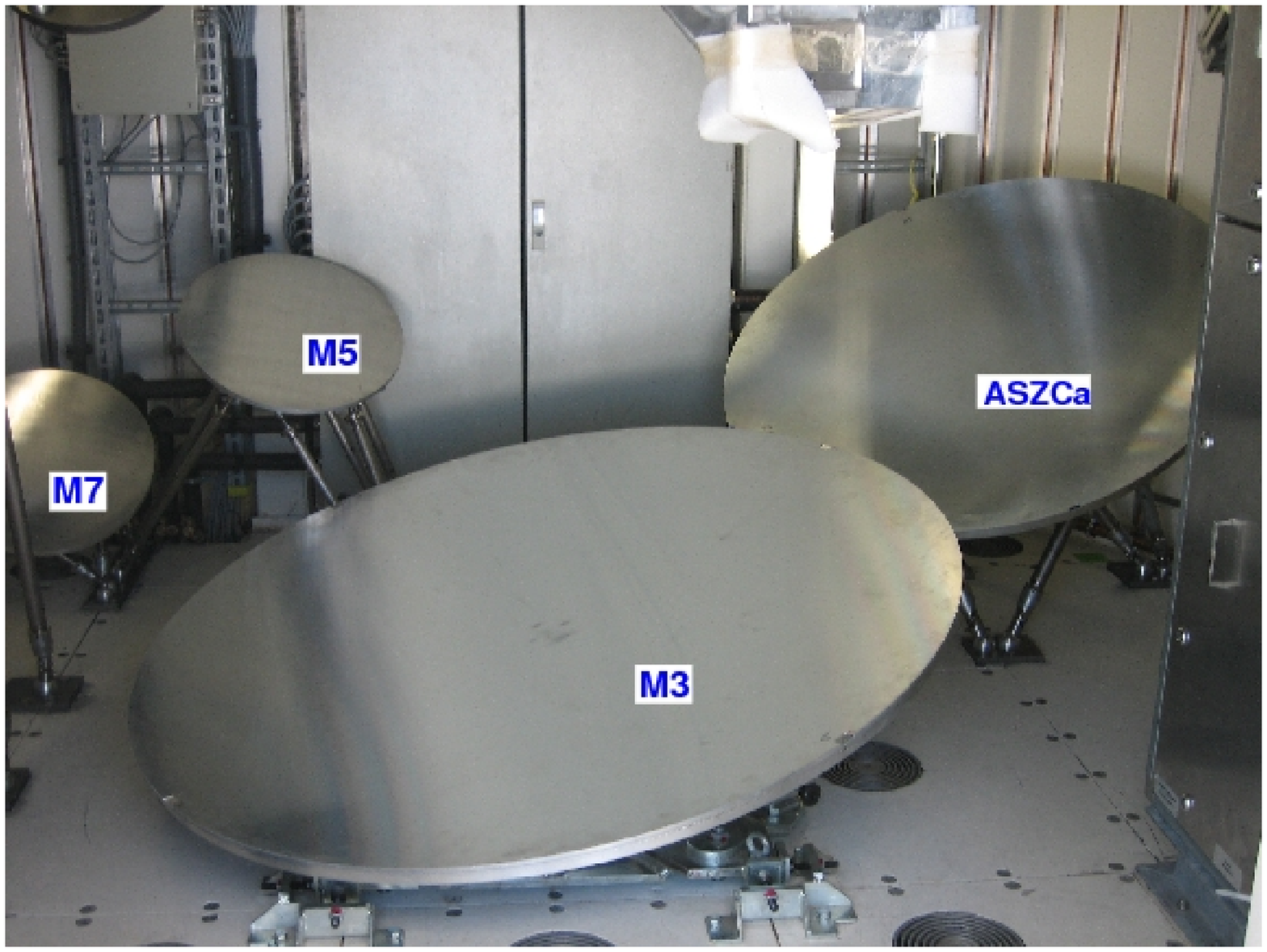}
\includegraphics[width=.95\linewidth, bb=20 20 560 740,clip]{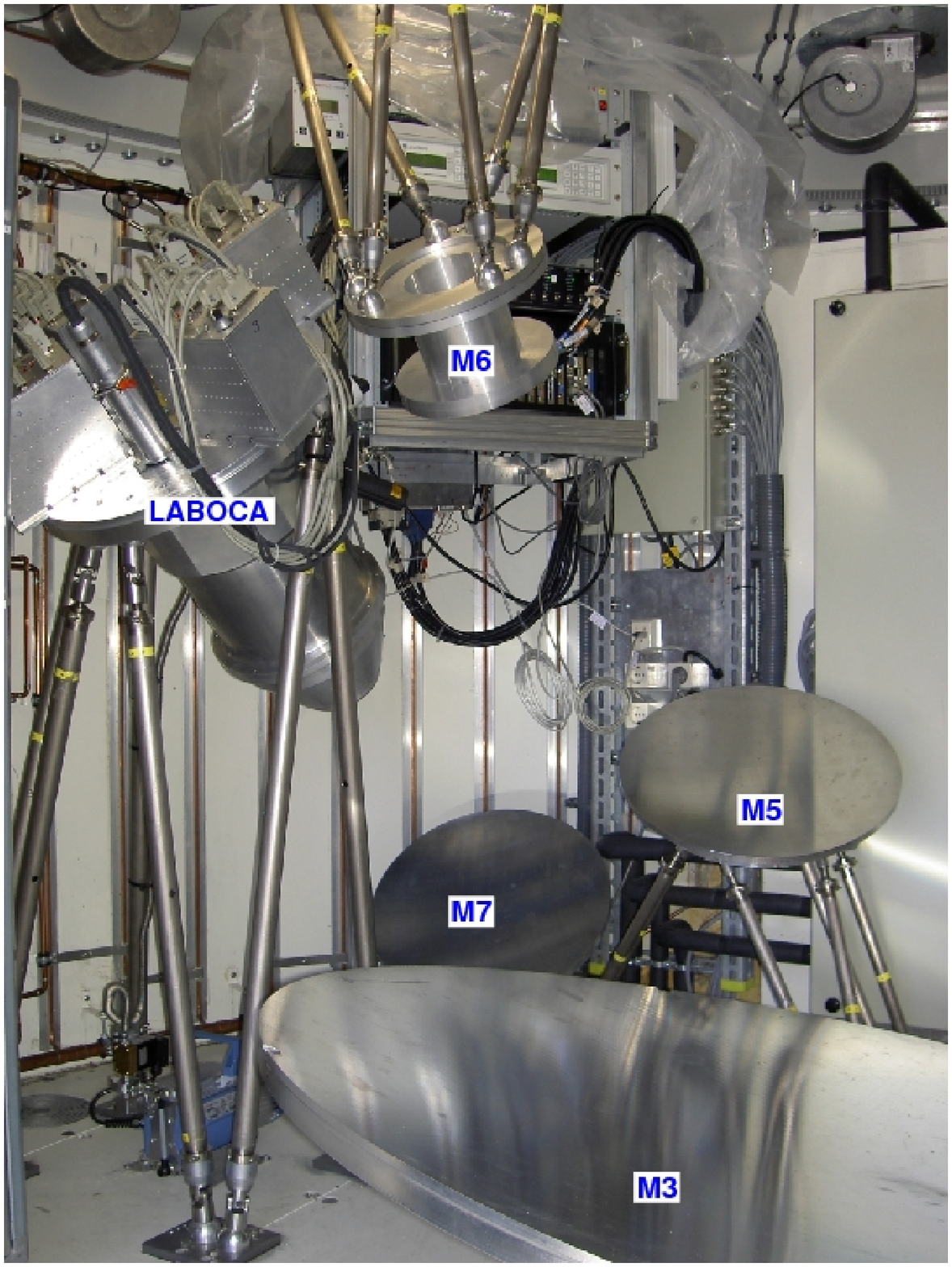} 
\caption
{
{\it Top:}~The mirrors affixed to the floor of the Cassegrain cabin of APEX.  Left to right: M7, M5, M3 and
one of the mirrors of the ASZCa experiment.  In this picture, mirror M3 is positioned for ASZCa.
{\it Bottom:}~The receiver, M3 in the position for LABOCA, M5, M6 and M7.
See also Fig.~\ref{fig:opticsall}.
}
\label{fig:receiver}
\end{figure}
\subsection{Manufacture} 
Mirror M3, an off-axis paraboloid, has been
manufactured by a machine shop at the Lawrence Berkeley National
Laboratory (LBNL, Berkeley, CA, USA) and is common to both LABOCA and the ASZCa
experiment.  M3 has a diameter of 1.6\,m and a surface accuracy of
18\,\micron  rms but LABOCA uses only the inner 80\,cm disk.  It is
attached to a bearing on the floor of the Cassegrain cabin, aligned to
the optical axis of APEX.  Operators can manually rotate the mirror in
order to direct the telescope beam to LABOCA or to ASZCa alternatively
(see Fig.~\ref{fig:receiver}).

Mirrors M4 and M6 are flat, have a diameter of 42\,cm and 26\,cm,
respectively (manufactured by Kugler\footnote{\tiny{\url{http://www.kugler-precision.com/}}}, Salem, Germany).
They are of optical quality and are both affixed to the ceiling of the cabin.
In a near future, mirror M6 will be replaced by the reflection-type half-wave
plate of the PolKa polarimeter \citep{2004A&A...422..751S}.

Mirrors M5 and M7 are off-axis aspherics, both 50\,cm in diameter, and
are affixed to the floor of the cabin. They have been designed and
manufactured at the MPIfR and have a surface accuracy of 7 and
5\,\micron  rms respectively.
\subsection{Installation and alignment} 
All the mirrors of LABOCA
(with exception of M3) and the receiver itself are mounted on hexapod positioners (see Fig.~\ref{fig:receiver})
provided by VERTEX Antennentechnik\footnote{\tiny{\url{http://www.vertexant.de/}}} (Duisburg, Germany).
Each hexapod is made of an octahedral assembly of struts and has six degrees of freedom (x, y, z, pitch,
roll and yaw).  The lengths of the six independent legs can be changed
to position and orient the platform on which the mirror is mounted.
VERTEX provided a software for calculating the required leg extensions
for a given position and orientation of the platforms.
A first geometrical alignment was performed during the first week of September
2006, using a double-beam laser on the optical axis of the telescope
and plane replacement mirrors in place of the two active mirrors M5
and M7. The alignment has been checked using the bolometers and hot
targets (made of absorbing\footnote{ECCOSORB AN, Emerson \& Cuming, Rundolph, MA, USA, \tiny{\url{http://www.eccosorb.com}}} material)
at different places along the beam, starting at the focal plane and
following the path through all the reflections up to the receiver's
window.  The alignment has been furthermore verified and improved in
February 2008.
%
%
\section{Cryogenics}\label{sec:cryo}
\subsection{Cryostat} 
The bolometer array of LABOCA is designed to be operated at a temperature
lower than 300\,mK.  This temperature is provided by a cryogenic
system made of a wet cryostat, using liquid nitrogen and liquid
helium, in combination with a two-stage sorption cooler.  A commercial
8-inch cryostat, built by Infrared Labs\footnote{\tiny{\url{http://www.irlabs.com/}}} (Tucson, AZ, USA), has been
customized at the MPIfR to accommodate the double-stage sorption
cooler, the bolometer array, cold optics and cold electronics.  A high
vacuum in the cryostat is provided by an integrated turbomolecular
pump backed by a diaphragm pump.  Operational vacuum is reached in one
single day of pumping.

The cryostat incorporates a 3-liter reservoir of liquid nitrogen and a
5-liter reservoir of liquid helium.  After producing high-vacuum
($\sim$10$^{-6}$\,mbar), the cryostat is filled with the liquid
cryogens.  The liquid nitrogen is used to provide thermal shielding at
77\,K in our labs in Bonn (standard air pressure, 1013\,mbar) and at
73.5\,K at APEX (5107\,m above the sea level) where the air pressure
is almost one half of the standard one (about 540\,mbar).

The liquid helium provides a thermal shielding at 4.2\,K at standard
pressure and 3.7\,K at the APEX site.  To keep it operational, the
system must be refilled once per day.  The refilling operation
requires about 20~minutes.
\subsection{Sorption cooler}
The cryostat incorporates a commercial two-stage closed-cycle sorption
cooler, model SoCool \citep{2002AIPC..613.1233D} manufactured by
Air-Liquide\footnote{\tiny{\url{http://www.airliquide.com/}}} (Sassenage,
France).  In this device, a \element[][4]{He} sorption cooler is used to
liquefy \element[][3]{He} gas in the adjacent, thermally coupled, \element[][3]{He}
cooler.  The condensed liquid \element[][3]{He} is then sorption pumped to
reach temperatures as low as 250\,mK, in the absence of a thermal
load.  Therefore, the double stage design makes it possible to cool the
bolometer array down to a temperature lower than 300\,mK starting from the
temperature of the liquid helium bath at atmospheric pressure. This
makes the maintenance of the system much simpler than that of other
systems, where pumping on the liquid helium bath is required.  The two
sorption coolers are closed systems, which means they do not require
any refilling of gas and can be operated from the outside of the
cryostat, simply by applying electrical power.

To keep the bolometers at operation temperature, the sorption cooler
needs to be recycled.  The recycling is done by application of a
sequence of voltages to the electric lines connected to thermal
switches and heaters integrated in the sorption cooler. A typical
recycling procedure requires about two hours and can be done manually
or in a fully automatic way controlled by the frontend computer (see
Sect.~\ref{sec:fepc}).  At the end of the recycling process, both
gases, \element[][4]{He} and \element[][3]{He}, have been liquefied and the controlled
evaporation of the two liquids provides a stable temperature for many
hours.  After the recycling of the sorption cooler, the bolometer
array reaches 285\,mK.  The hold time of the cooler, usually between
10 and 12 hours, strongly depends on the parameters used during the
recycling procedure.
The end temperature is a function of elevation and can be affected by telescope movements,
leading to temperature fluctuations $\la500\,$\microK within one scan,
during regular observations, and $\la3\,$mK for wide elevation turns
(e.g. during skydips, see Sect.~\ref{sec:skydips}).
\subsection{Temperature monitor}\label{sec:thermonitor}
The cryostat of LABOCA incorporates 8
thermometers to measure the temperature at the different stages:
liquid nitrogen, liquid helium, the two sorption pumps, the two
thermal switches, evaporator of the \element[][4]{He} and evaporator of the
\element[][3]{He}. Two LS218\footnote{\tiny{\url{http://www.lakeshore.com/temp/mn/218po.html}}}
devices (Lake Shore Inc., Westerville, OH, USA) are used to monitor the thermometers
and to apply the individual temperature calibrations in real-time. 
The temperature of the \element[][3]{He} stage is measured with higher accuracy with the use of a
resistance bridge AVS-47\footnote{\tiny{\url{http://www.picowatt.fi/avs47/avs47.html}}}
(Picowatt, Vantaa, Finland), with an error of $\pm$5\,\microK.
Control and monitor of the cryogenic
equipment can be done remotely via the frontend computer (see
Sect.~\ref{sec:fepc}).
\begin{figure}[t]
\centering
\includegraphics[width=.98\linewidth, bb=124 93 764 523, clip]{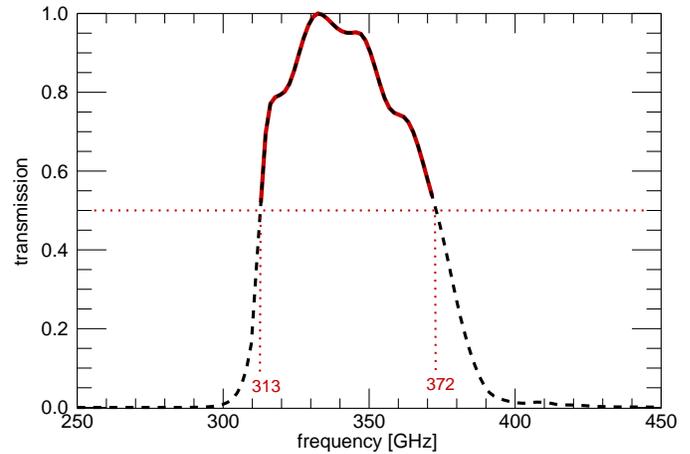}
\caption
{
Spectral response of LABOCA, relative to the maximum.
The central frequency is 345\,GHz, the portion with 50\% or more
transmission is between 313 and 372\,GHz.
}
\label{fig:passband}
\end{figure}
%
\section{Cold optics}\label{sec:coptics}
\subsection{Passband definition}\label{sec:passband}
Inside the cryostat, a set of cold filters, mounted on
the liquid nitrogen and liquid helium shields define the spectral
passband, centered at a wavelength of 870\,\micron  (345\,GHz) and about
150\,\micron  (60\,GHz) wide (see Fig.~\ref{fig:passband}).  The filters
have been designed and manufactured at MPIfR in collaboration with the
group of V.~Hansen (Theoretische Elektrotechnik\footnote{\tiny{\url{http://www.tet.uni-wuppertal.de/}}}, Bergische
Universit\"at Wuppertal, Germany) who provided theoretical support and
electromagnetic simulations.  The passband is formed by an
interference filter made of inductive and capacitive meshes embedded
in polypropylene.  The low frequency edge of the band is defined by
the cut-off of the cylindrical waveguide of each horn antenna (see
also Sect.~\ref{sec:horns}).  A freestanding inductive mesh behind the
window-lens provides shielding against radio interference.
\subsection{Horn array}\label{sec:horns} 
A monolithic array of conical
horn antennas, placed in front of the bolometer wafer, concentrates
the radiation onto the bolometers.  295 conical horns have been
machined into a single aluminum block by the MPIfR machine shop. In
combination with the tertiary optics, the horn antennas are optimized
for coupling to the telescope's main beam at a wavelength of
870\,\micron . The grid constant of the hexagonal array is
4.00\,mm. Each horn antenna feeds into a circular wave guide with a
diameter of 0.54\,mm, acting as a high-pass filter.
\section{Detector}\label{sec:detector}
\begin{figure*}[t]
\centering
\includegraphics[width=.32\linewidth, bb= 0 0 200 200, clip]{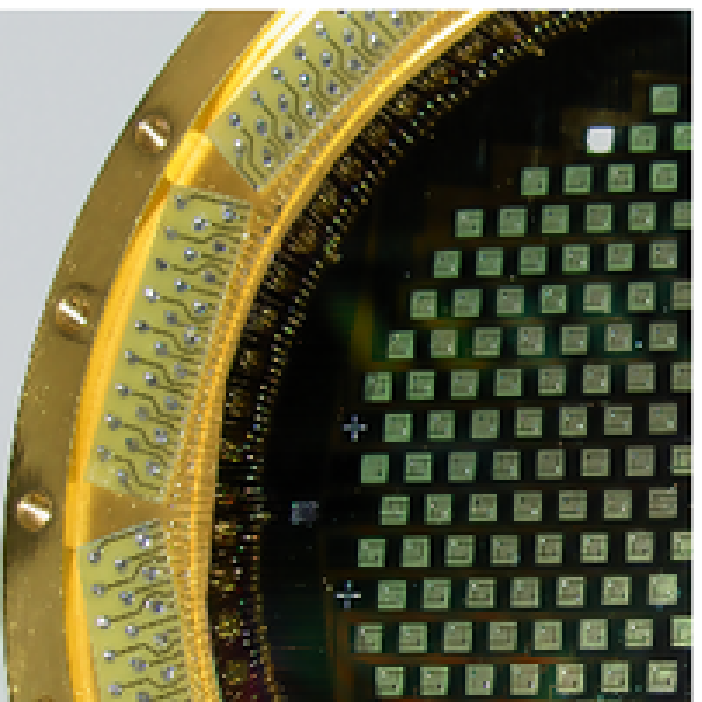}
\includegraphics[width=.32\linewidth, bb= 0 0 200 200, clip]{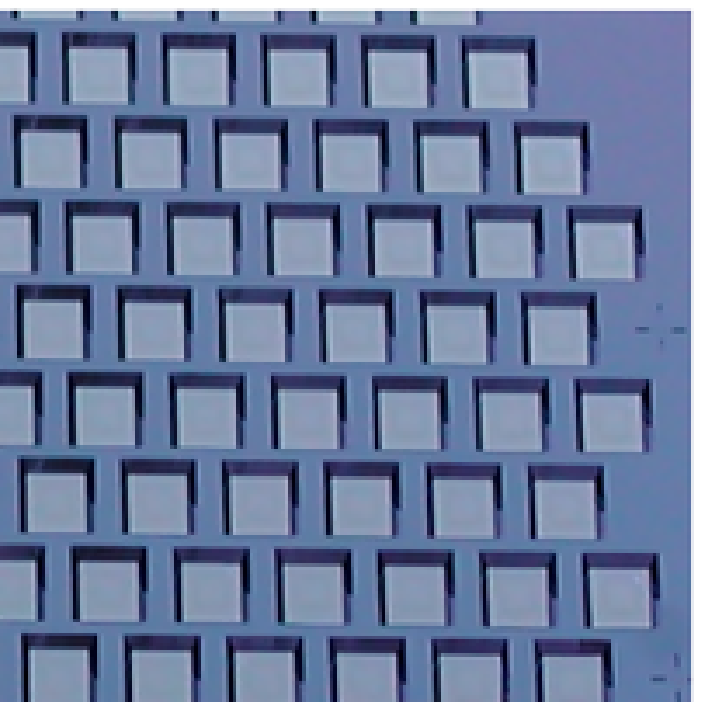}
\includegraphics[width=.32\linewidth, bb= 0 0 200 200, clip]{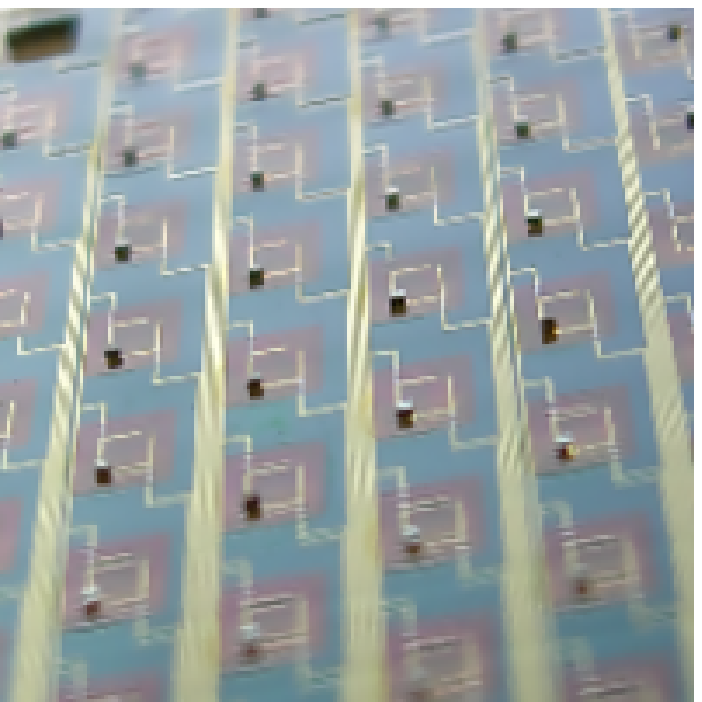}
\caption
{
Pictures of the bolometer array of LABOCA.
{\it Left:}~A detail of the array mounted in its copper ring. Some bonding wires are visible.
{\it Center:}~The side of the array where the bolometer cavities are etched in the silicon wafer.
{\it Right:}~Wiring side of the array. The thermistors are visible as small cubes on the membranes.
One broken membrane is visible on the top left corner. See also Fig.~\ref{fig:nakedarray}.
}
\label{fig:arraypics}
\end{figure*}
\subsection{Array design and manufacture} \label{sec:array}
The bolometer array of LABOCA is made of 295 composite bolometers arranged in an hexagonal
layout consisting of a center channel and 9 concentric hexagons (see Fig.~\ref{fig:nakedarray}
and Fig.~\ref{fig:arraypics}). The array is manufactured on a 4-inch silicon wafer coated on both sides with
a silicon-nitride film by thermal chemical vapor deposition.  On one side of the wafer,
295 squares are structured into the silicon-nitride film used as a mask for the alkaline KOH
etching of the silicon, producing freestanding, unstructured silicon-nitride membranes, only
400\,nm thick (see the center picture in Fig.~\ref{fig:arraypics}).
On the other side, the wiring is created by microlithography of
niobium and gold thin metal layers.  The bolometer array is mounted
inside a gold-coated copper ring and is supported by about 360 gold
bonding thin wires (see Fig.~\ref{fig:arraypics},~left), providing the
required electrical and thermal connection. This copper ring
also serves as a mount for the backshort reflector, at $\lambda$/4 distance from
the array, and 12 printed circuit boards hosting the load resistors
and the first electronic circuitry (see Sect.~\ref{sec:jfet}).
\begin{figure}[t]
\centering
\includegraphics[width=.94\linewidth, bb=0 0 233 173, clip]{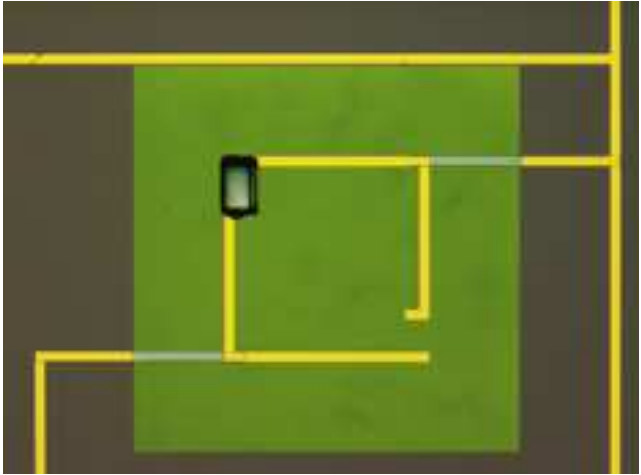}
\caption
{
Wiring side of one bolometer of LABOCA, seen under a microscope.
The membrane is the green colored area. The black box is the NTD thermistor.
The wiring is made of gold (yellow) and niobium (gray) thin metal layers.
See also Fig. \ref{fig:arraypics}.
}
\label{fig:bolo}
\end{figure}
\subsection{Bolometer design and manufacture} 
The description of a composite bolometer can be simplified as the combination of two elements:
an extremely sensitive temperature sensor, called thermistor, and a radiation absorber. 
In the bolometers of LABOCA, the absorbing element is made of a thin film of titanium deposited on
the unstructured silicon-nitride membranes.  LABOCA uses neutron-transmutation doped germanium semiconducting chips
\cite[NTD,][]{1982STIN...8328412H} as thermistors.  The NTD thermistors are made from ultra-pure germanium doped by
neutron-transmutation in a nuclear reactor.  The thermistors, needed to measure the temperature of the absorber,
are soldered on the bolometer membranes to gold pads connected to the outer edge of the silicon wafer through a
pair of patterned niobium wires, superconducting at the operation temperature (see Fig.~\ref{fig:bolo} and Fig. \ref{fig:arraypics}, right).
The soldering of the thermistors was the only manual step in the manufacture of the bolometers.
LABOCA has so-called {\it flatpack} NTD thermistors, which have two ion
implanted and metalized contacts on one side of the thermistor block.
The chips are optimized to work at a temperature lower than 300\,mK,
where they show an electric impedance in the range of 1 to 10\,MOhms.
%
\section{Cold electronics}\label{sec:jfet} 
\subsection{Bias resistors}
Given the high impedance of the NTD thermistors, the electric scheme
of the first bolometer circuitry requires very high impedance load
resistors, which are needed to current bias the bolometers.  These
bias resistors are 312 identical 30\,MOhm chips, made of a nichrome
thin film deposited on silicon substrate (model MSHR-4 produced by
Mini-Systems Inc.\footnote{\tiny{\url{http://www.mini-systemsinc.com/}}},
N. Attleboro, MA, USA) mounted on 12 identical printed circuit boards, to
form 12 groups of 26 resistors. This configuration is reflected in the
following distribution of the bolometer signals.  The circuit boards
are mounted on the same copper ring which holds the bolometer array
(see Sect.~\ref{sec:array})
and electrically connected to the bolometers through miniature RF
filters.

\subsection{Junction Field Effect Transistors (JFETs) source followers} 
The high impedance of the bolometers makes the system
sensitive to microphonic noise pickup, therefore JFETs (Toshiba 2SK369)
are used as source followers in order to decrease the impedance of the
electric lines down to a few kOhms before they reach the high gain
amplification units at room temperature.  Following the wiring scheme
of the bias resistors, the JFETs are also in groups of 26 soldered
onto 12 printed circuit boards, electrically connected to the
corresponding bias resistors by 12 flat cables made of manganin traces embedded in
Kapton\footnote{DuPont, \tiny{\url{http://www2.dupont.com/Kapton/en_US/}}}
(manufactured by VAAS Leiterplattentechnologie\footnote{\tiny{\url{http://www.vaas-lt.de/}}},
Schw\"abisch-Gm\"und, Germany), thermally shunted to the liquid helium tank.
The 12 JFET boards are assembled in groups of
three into four gold-coated copper boxes, thermally connected to the
liquid nitrogen tank.  Inside each box, during regular operation, the
78 JFETs are self-heated to a temperature of about 110\,K, where they
show a minimum of their intrinsic noise.  Through the connections in
the JFET boxes, the wiring scheme of 12 groups of 26 channels is
translated to a new scheme of 4 groups of 80 channels.
%
%
\section{Warm electronics and signal processing}\label{sec:electronics}
\subsection{Amplifiers}\label{sec:ampli}
The 312 channels exiting the
cryostat of LABOCA are distributed to 4 identical, custom made,
amplification units, providing 80 channels each. Of these, 295 are
bolometers, 17 are connected to 1\,MOhm resistors mounted on the
bolometer ring (used for technical purposes, like noise monitoring and
calibrations) and the remaining 8 are not connected. Each
amplification unit is made of 16 identical printed circuit boards and
each board provides 5 low noise, high gain amplifiers. Each unit also
includes a low noise battery used to generate the bias voltage and the
circuitry to produce the AC biasing and perform real time demodulation
of the 320 signals. The AC bias reference frequency is not internally
generated but is provided from the outside, thus allowing
synchronization of the biasing to an external frequency source.

Each amplification unit is equipped with a digital interface controlled by a microprocessor programmed to provide remote control of
the amplification gain and of the DC offset removal procedure. This is required because LABOCA is a total power receiver and the
signals carry a floating DC offset which could exceed the dynamic range of the data acquisition system.
To avoid saturation, therefore, the DC offsets are measured and subtracted from the signals at the beginning of every integration.
The values of the 320 offsets are temporarily stored in a local memory and, at the end of the observation, are written into the
corresponding data file for use in the data reduction process.
The digital lines use the I$^2$C protocol\footnote{Inter-Integrated Circuit, a serial bus to connect hardware devices.}
and are accessible remotely via the local network through I$^2$C-to-RS232\footnote{Recommended Standard 232, a standard for serial
binary data communication.} interfaces controlled by the frontend computer (see Sect.~\ref{sec:fepc}).
The amplification gain can be set in the range from 270 to 17280.
\subsection{Data acquisition}\label{sec:daq}
The 320 output signals from the 4 amplification units are digitized over 16 bits by 4 multifunction data acquisition (DAQ) boards
(National Instruments\footnote{\tiny{\url{http://www.ni.com/}}} M-6225-PCI),
providing 80 analog inputs each and synchronized to the same sample
clock by a RTSI\footnote{Real-Time System Integration, a bus used to share and exchange
timing and control signals between multiple boards.} bus . The maximum data
sampling is 2500\,Hz and the dynamic range can be selected over 5
predefined ranges. The four boards provide also 24 digital
input/output lines each, some of them used for the generation of the
bias reference frequency and to monitor the digital reference signals
(sync/blank) of the wobbler.  For the time synchronization of the data
to the APEX control software \citep[APECS,][]{2006A&A...454L..25M} the
data acquisition system is equipped with a precision time interface
(PCI-SyncClock32\footnote{\tiny{\url{http://www.brandywinecomm.com}}} from Brandywine Communications, Tustin, CA, USA)
synchronized to the station GPS clock via
IRIG-B\footnote{Inter Range Instrumentation Group, standardized time code format.} time code signal.

The AC bias reference frequency is provided by the data acquisition
system as a submultiple of the sampling frequency, thus synchronizing
the bias to the data sampling.  Typical values used for observations
are 1\,kHz for the sampling rate and 333\,Hz for the AC bias.  The
backend computer has two network adapters: one is connected to the
local area network, the other one is exclusively used for the output
data stream and is connected in a private direct network with
the bridge computer (see also Sect.~\ref{sec:bridge}).

The data acquisition software is entirely written using
LabVIEW\footnote{\tiny{\url{http://www.ni.com/labview/}}} (National Instruments).
The drivers for the data acquisition hardware are provided by the NI-DAQmx\footnote{\tiny{\url{http://www.ni.com/dataacquisition/nidaqmx.htm}}} package.
Custom drivers for LabVIEW have been developed to access the GPS clock interface.
The backend software runs a server to stream the output data to the bridge computer and allows remote control
and monitoring of the operation via a CORBA\footnote{Common Object Request Broker Architecture, a set of
standards which define the protocol for interaction between the objects of a distributed system} object interfaced
to the APECS through the local area network.
\subsection{Anti-aliasing filtering and downsampling}\label{sec:bridge}
The amplification units of LABOCA use the bias signal, which is a square waveform,
as reference to operate real-time demodulation of the AC-biased bolometer signals.
Therefore, all the frequencies present in the bolometer readout lines end up aliased
around the odd-numbered harmonics of the bias frequency. Microphonics pickup by the
high-impedance bolometers at a few resonant frequencies can produce a forest of lines in
the final readout, polluting even the lower part of the post-detection frequency band,
where the astronomical signals are expected.
\begin{figure}[t]
\centering
\includegraphics[width=.96\linewidth, bb=0 0 726 551,clip]{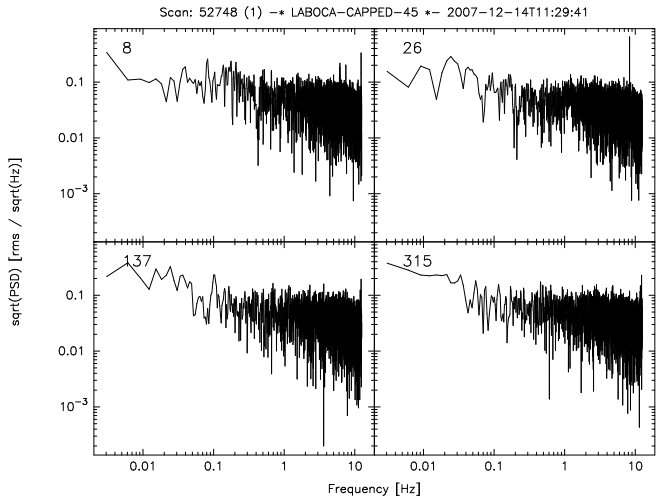}
\caption
{
Noise spectra for selected bolometers on a room temperature absorber (capped optics). 
The signals, sampled at 1\,kHz, were downsampled in real-time to 25\,Hz by the bridge computer.
The spectra are free of microphonic pick-ups and show the $1/f$ noise onset at $\sim$0.1\,Hz.
}
\label{fig:noisespectra}
\end{figure}
To overcome this, we introduced an intermediate stage into the
sampling scheme, the so-called {\it bridge computer}. The bolometer
signals, acquired by the backend at a relatively high sampling rate
(usually 1\,kHz), well above the rolling-off of the anti-alias filters
embedded in the amplifiers, are sent to the bridge computer where they
are digitally low-pass filtered and then downsampled to a much lower
rate (usually 25\,Hz), more appropriate for the astronomical signals
produced at the typical scanning speeds (see also
Sect.~\ref{sec:obsmodes}).  The digital real-time anti-alias filtering
and downsampling is performed by a non-recursive convolution filter
with a Nutall window such that its rejection at the Nyquist frequency
is $\sim$3\,dB and falls steeply to $\sim$100\,dB soon beyond that.
The bias reference frequency was accordingly selected to maximize the
astronomically useful post-detection bandwidth.  The resulting
bolometer signals are generally white between 0.1--12.5\,Hz and free
of unwanted microphonic interference (see Fig.~\ref{fig:noisespectra}).

To seamlessly integrate the bridged readout into the APEX control
system, the bridge computer also acts as a fully functional virtual
backend that forwards all communication between the actual backend
computer and the control system, while intercepting and reinterpreting
any commands of interest for the downsampling scheme.
\subsection{Frontend computer}\label{sec:fepc}
The so-called {\it frontend computer} communicates with the hardware
of LABOCA through the local area network. It is devoted to monitor and
control most of the electronics of the instrument (e.g.~monitoring of
all the temperature stages, control of the sorption cooler,
calibration unit, power lines\dots) and also provides a CORBA object
for interfacing to APECS, allowing remote operation of the system.
The frontend software is entirely written using LabVIEW and custom
drivers have been developed for some hardware devices embedded in LABOCA.
%
%
%
\begin{figure}[t]
\centering
\includegraphics[width=.94\linewidth, bb=33 3 259 420, clip]{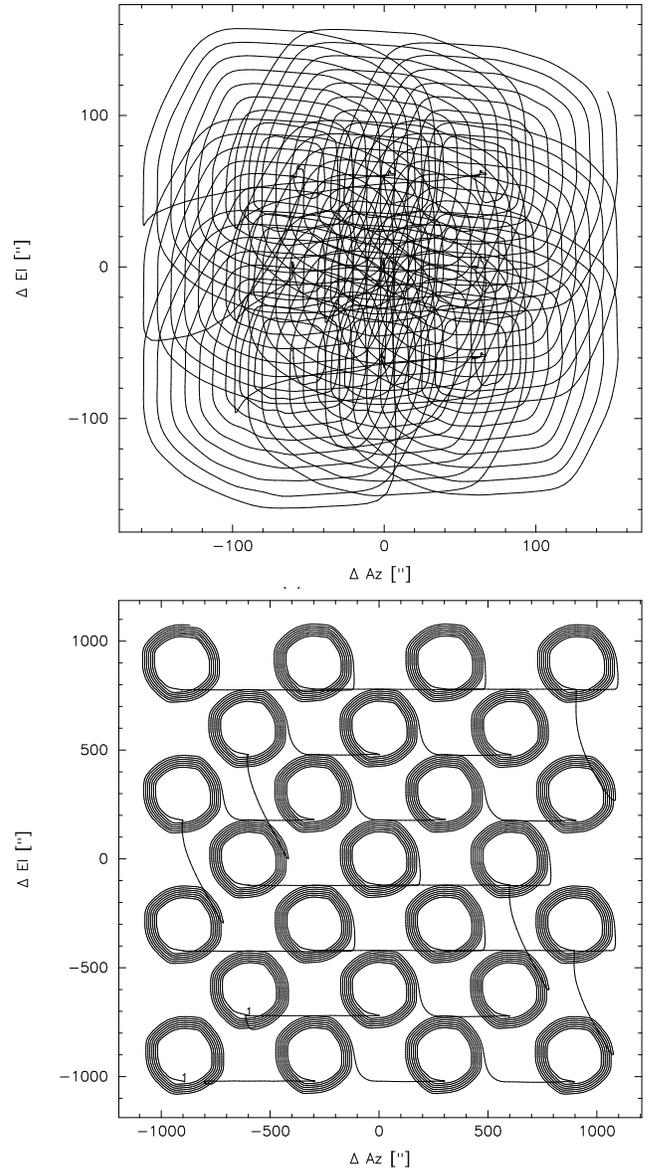}
\caption{
Examples of raster-spiral patterns. The two plots show the scanning pattern of the central beam
of the array in horizontal coordinates.  The compact pattern shown in
the top panel is optimized to map the field of view of LABOCA.
The large scale map shown in the bottom panel consists of 25 raster
positions and covers a field of $0.5\times0.5$ degrees.  The complete
scan required 19 minutes and the final map covers about
$2000\times2000$ arcseconds with uniform residual rms noise.
}
\label{fig:spiralrast}
\end{figure}

\section{Observing modes}\label{sec:obsmodes} 
\subsection{Mapping modes}\label{sec:mapmodes}
In order to reach the best signal-to-noise ratio using the fast scanning technique \citep{2001A&A...379..735R}
with LABOCA, the frequencies of the signal produced by scanning across the source need to match the white noise part of
the post-detection frequency band (0.1--12.5\,Hz, see Fig.~\ref{fig:noisespectra}), mostly above the frequencies of the
atmospheric fluctuations. Thus, with the $\sim$19$''$~beam (see Sect.~\ref{sec:beam}),
the maximum practical telescope scanning speed for LABOCA is about $4'$/s.
This is also the limiting value to guarantee the required accuracy in the telescope position information of each sample.
The minimum scanning speed required for a sufficient source modulation depends on the atmospheric stability and
on the source structure and is typically about $30''$/s.
The APEX control system currently supports two basic scanning modes: on-the-fly maps (OTF) and spiral scanning patterns.
\subsubsection{Spirals}
Spirals are done with a constant angular speed and an increasing radius, therefore the linear scanning velocity
is not constant but increases with time. We have selected two spiral modes of 20\,s and 35\,s integration time,
both producing fully sampled maps of the whole FoV with scanning velocities limited between $1'$/s and $2\rlap{.}'5$/s.
The spiral patterns are kept compact (maximum radius $\lesssim2'$), the scanned area on the sky is only slightly
larger than the FoV and most of the integration time is spent on the central $11'$ of the array.
These spirals are the preferred observing modes for pointing scans on sources with flux densities down to a few Jy.

For fainter sources, the basic spiral pattern can be combined with a raster mapping mode (raster-spirals) on a grid
of pointing positions resulting in an even denser sampling of the maps and longer integration time (see Fig.~\ref{fig:spiralrast}, top panel).
These compact raster-spirals give excellent results for sources smaller than the FoV of LABOCA
and are also suitable for integrations of very faint sources.

The flexibility in the choice of the spiral parameters also allows spiral
observing patterns to be used to map fields much larger than the FoV.
The bottom panel of Fig.~\ref{fig:spiralrast} shows an example of
raster of spirals optimized to give an homogeneous coverage across
a field of $0.5\times0.5$ degrees. In this case, the spirals start
with a large radius and follows an almost circular scanning pattern
for each raster position.  This mapping mode is very useful for
cosmological deep field surveys since co-adding several such
raster-spiral scans, taken at different times and thus at different
orientations, provides an optimal compromise between telescope
overheads, uniform coverage and cross-linking of individual map
positions \citep[see ][]{Kovacs2008}.

\subsubsection{On-the-fly maps (OTF)}
OTF scans are rectangular scanning patterns produced moving
back-and-forth along alternating rows with a linear constant speed and
accelerating only at the turnarounds.  They can be performed in
horizontal or equatorial coordinates and the scanning direction can be
rotated relatively to the base system for both coordinate systems. OTF
patterns have been tested for maps on the scales of the FoV up to long
slews across the plane of the Milky Way (2 degrees).  Small
cross-linked OTFs (of size $\sim$FoV of LABOCA) give results
comparable to the raster-spirals \citep{Kovacs2008}, but the
overheads are much larger at a scanning speed of $2'$/s. For larger
OTFs the relative overheads decrease.
\subsection{Ancillary modes}
\subsubsection{Point}
The standard pointing procedure consists of one subscan in spiral
observing mode and results in a fully sampled map of the FoV of
LABOCA. The pointing offsets relative to the pointing model are
computed via a two-dimensional Gaussian fit to the source position in
the map using a BoA pipeline script (see Sect.~\ref{sec:redu}). Note
that this pointing procedure is not limited to pointing scans of the
central channel of the array but works independently of the reference
bolometer, thus allowing pointing scans centered on the most sensitive
part of the array.
\subsubsection{Focus}
The default focusing procedure is made of 10 subscans at 5 different
subreflector positions and 5 seconds of integration time each.  This
is the only observing mode without scanning telescope motion.
As a result, we are currently restricted to sources brighter than
the atmospheric variations (Mars, Venus, Saturn and Jupiter).
However, initial tests confirmed that using the wobbler to modulate the
source signal allows focusing on weaker sources, too.
This is the only observing mode, so far, for which the use of the wobbler with LABOCA has been tested.
\subsubsection{Skydips}\label{sec:skydips}
The attenuation of the astronomical signals due to the atmospheric
opacity is determined with skydips. These scans measure the power of
the atmospheric emission as a function of the airmass while tipping
the telescope from high to low elevation. A skydip procedure consists
of two steps: a hot-sky calibration scan, to provide an absolute
measurement of the sky temperature, followed by a continuous tip scan
in elevation; see also Sect.~\ref{sec:opacity}.
\section{Performance on the sky and sensitivity}\label{sec:performance}
\subsection{Number of channels}\label{sec:numchans}
\begin{figure}[t]
\centering
\includegraphics[width=.98\linewidth, bb=65 65 535 525, clip, angle=-90]{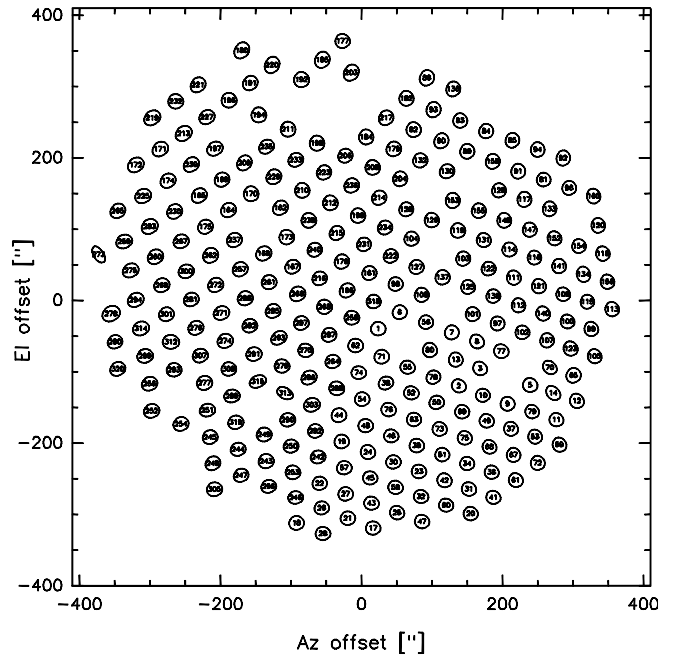}
\caption{
Footprint of LABOCA on sky, measured with a beam map on the planet Mars.
The ellipses represent the FWHM shape of each beam on sky, as given by a
two-dimensional Gaussian fit to the single-channel map of each
bolometer.  Only bolometers with useful signal-to-noise ratios are
shown in this map. See also Fig.~\ref{fig:sensitivity}.
}
\label{fig:beams}
\end{figure}
At the time of the commissioning, the number of channels with sky
response was 266 (90\% of the nominal 295 bolometers).  Of these, 6
channels show cross-talk and 12 channels have low sensitivity (less
than 10\% of the mean responsivity).  Two additional bolometers have
been blinded by blocking their horn antennas with absorber material so
they can be used to monitor the temperature fluctuations of the array.
The number of channels used for astronomical observations is therefore
248 (84\% yield, see Fig.~\ref{fig:beams}).
\subsection{Array parameters}
Position on sky and relative gain of each bolometer are derived from
fully sampled maps (hereafter called {\it beam maps}) of the planets
Mars and Saturn (see Fig.~\ref{fig:beams}), besides giving a realistic
picture of the optical distortions over the FoV.  Variations among
maps were found to be within a few arc seconds for the positions and
below 10\% for peak flux densities.  A table with average receiver
parameters (RCP)\footnote{The latest RCP table is available
at:\\
\tiny{\url{http://www.apex-telescope.org/bolometer/laboca/calibration/LABOCA-centred.rcp}}}
is periodically computed from beam maps and implemented in the BoA software (see Sect.~\ref{sec:redu}).
The accuracy of the relative bolometer positions from this master RCP is
typically below $1''$ (5\% of the beam size) and the gain accuracy is
better than 5\%, confirming the good quality (small distortion over
the entire FoV) of the tertiary optics (see also
Sect.~\ref{sec:optics}).
\subsection{Beam shape}\label{sec:beam}
\begin{figure}[t] 
\centering
\includegraphics[width=.98\linewidth, bb=0 0 486 453, clip]{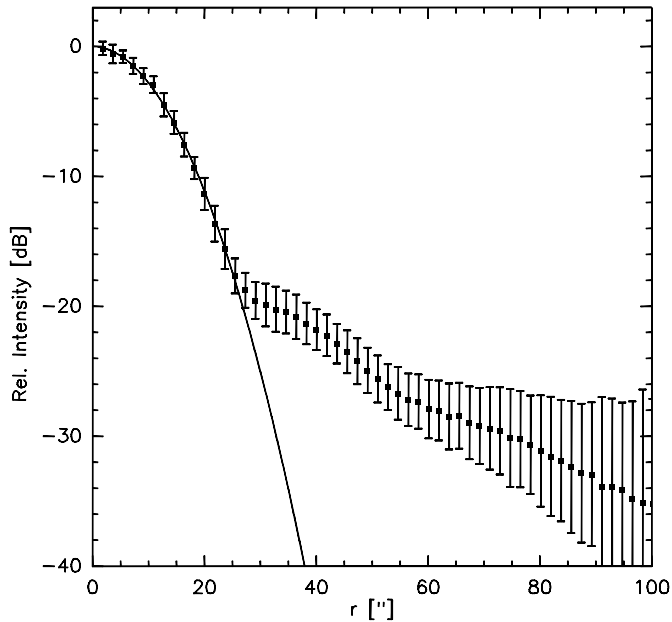} 
\caption
{
Radial profile of the LABOCA beam derived by averaging the beams of all 248 functional bolometers
from fully sampled maps on Mars.  The error bars show the standard deviation.
The main beam is well described by a Gaussian with a FWHM of $19\rlap{.}''2\pm0\rlap{.}''7$
and starts to deviate at $-$20\,dB.} 
\label{fig:beamshape}
\end{figure}
The LABOCA beam shape was derived for individual bolometers from fully
sampled maps on Mars (see Fig.~\ref{fig:beams}) as well as on pointing
scans on Uranus and Neptune.  Both methods lead to comparable results
and give an almost circular Gaussian with a FWHM of
$19\farcs2\pm0\farcs7$ (after deconvolution from the source
and pixel sizes).  We also investigated the error beam pattern of LABOCA
on beam maps on Mars and Saturn (see Fig.~\ref{fig:beamshape}).  The
beam starts to deviate from a Gaussian at $-$20~dB (1\% of the peak
intensity).  The first error beam pattern can be approximated by a
Gaussian with a peak of $-18.3$~dB and a FWHM of~$70''\pm5''$, the
support legs of the subreflector are visible at the $-25$~dB (0.3\%
level). The fraction of the power in the first error beam is $\sim$18\%.
\subsection{Calibration}\label{sec:calibra}
The astronomical calibration was achieved on Mars, Neptune and Uranus
and a constant conversion factor of $6.3\pm0.5$\,Jy\,beam$^{-1}$\,\microV$^{-1}$ was
determined between LABOCA response and flux density.  For the determination
of the calibration factor we have used the fluxes of planets determined with
the software ASTRO\footnote{GILDAS package, \tiny{\url{http://www.iram.fr/IRAMFR/GILDAS}}}. 
The overall calibration accuracy for LABOCA is about 10\%. 
We have also defined a list of secondary calibrators
and calibrated them against the planets (see Appendix~\ref{sec:appendix}).
In order to improve the absolute flux determination, the calibrators
are observed routinely every $\sim$2~hours between observations of
scientific targets.
\subsection{Sky opacity determination}\label{sec:opacity}
Atmospheric absorption in the passband of LABOCA attenuates the astronomical signals as $\exp(-\tau_{\rm los})$
where the line of sight optical depth $\tau_{\rm los}$ can be as high as 1 for observations at low elevation
with 2\,mm of precipitable water vapor (PWV), typical limit for observations with LABOCA.
The accuracy of the absolute calibration, therefore, strongly depends on the precision in the determination of $\tau_{\rm los}$.

We use two independent methods for determining $\tau_{\rm los}$. 
The first one relies on the PWV level measured every minute by the APEX radiometer broadly along the line of sight.
The PWV is converted into $\tau_{\rm los}$ using an atmospheric transmission model \citep[ATM,][]{2001ITAP...49.1683P} and
the passband of LABOCA (see Sect.~\ref{sec:passband}). The accuracy of this approach is limited by the knowledge of the passband,
the applicability of the ATM and the accuracy of the radiometer.

The second method uses skydips (see Sect.~\ref{sec:skydips}).
As the telescope moves from high to low elevation, $\tau_{\rm los}$ increases with airmass.
The increasing atmospheric load produces an increasing total power signal converted
to effective sky temperature ($T_{\rm eff}$) by direct comparison with a reference hot load.
The dependence of $T_{\rm eff}$ on elevation is then fitted to determine the zenith opacity~$\tau$,
used as parameter \citep[see, e.g.,][]{2004MNRAS.354..621C}.

Since LABOCA is installed in the Cassegrain cabin of APEX, when performing skydips the receiver suffers a wide,
continuous rotation (about 70 degrees in 20 seconds), which affects the stability of the sorption cooler,
thus inducing small variations of the bolometers temperature ($\sim$1--2\,mK).
These temperature fluctuations mimic an additional total power signal with amplitude comparable to the atmospheric signal.
The bolometers temperature, however, is monitored with high accuracy by the~\element[][3]{He}-stage thermometer (see Sect.~\ref{sec:thermonitor})
and by the two blind bolometers (see Sect.~\ref{sec:numchans}), making possible a correction of the skydip data.

The values of $\tau$ resulting from the skydip analysis are robust,
yet up to $\sim$30\% higher than those obtained from the radiometer.
There are several possible explanation for the discrepancy:
it could be the result of some incorrect assumptions going into the skydip model (e.g.~the sky temperature),
or the model itself may be incomplete. The non-linearity of the bolometers can be another factor.
The detector responsivities are expected to change with the optical load as $\sim\exp(-\gamma\,\tau_{\rm los})$ (to first order in $\tau$),
where $\gamma$ can be related to the bolometer constants \citep{Mather1984} and the optical configuration.
Combined with the sky response, the bolometer non-linearities would increase the effective skydip $\tau$ by a factor $(1+\gamma)$.

Our practical approach to reconciling the results obtained with the two methods has been to use a linear combination
of radiometer and skydip values such that it gives the most consistent calibrator fluxes
at all elevations\footnote{A text file (BoA readable, see Sect.~\ref{sec:redu}) containing the zenith opacities
calculated from the skydips and the radiometer is available at:\\
\tiny{\url{http://www.apex-telescope.org/bolometer/laboca/calibration/opacity/}}}.
The excellent calibration accuracy of LABOCA (see Appendix~\ref{sec:appendix}) underscores this approach.
\subsection{Sensitivity}\label{sec:sensitivity}
\begin{figure}[t] 
\centering 
\includegraphics[width=.98\linewidth, bb=0 -30 698 704, clip]{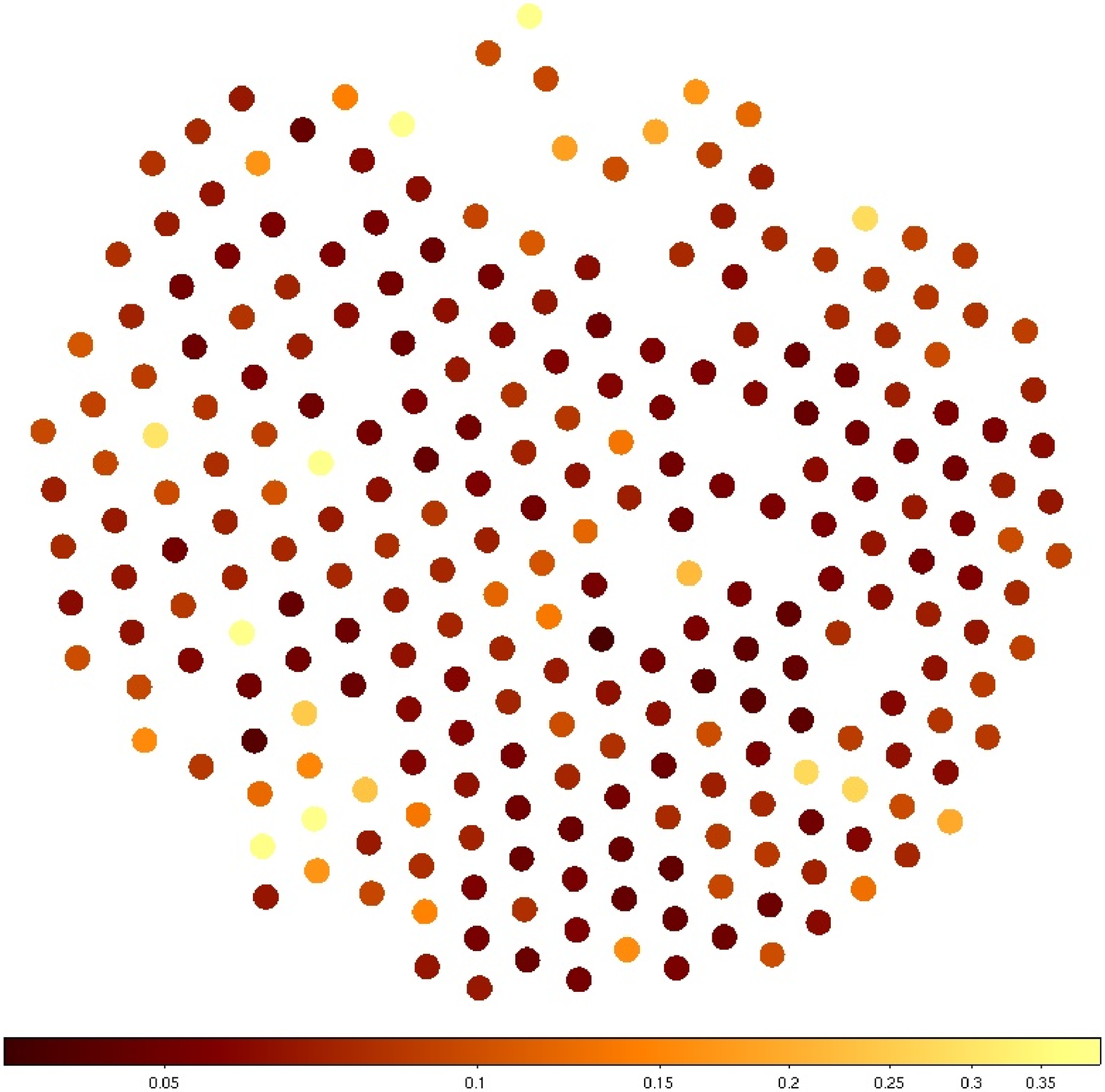}
\includegraphics[width=.98\linewidth, bb= 108 57 348 296,clip]{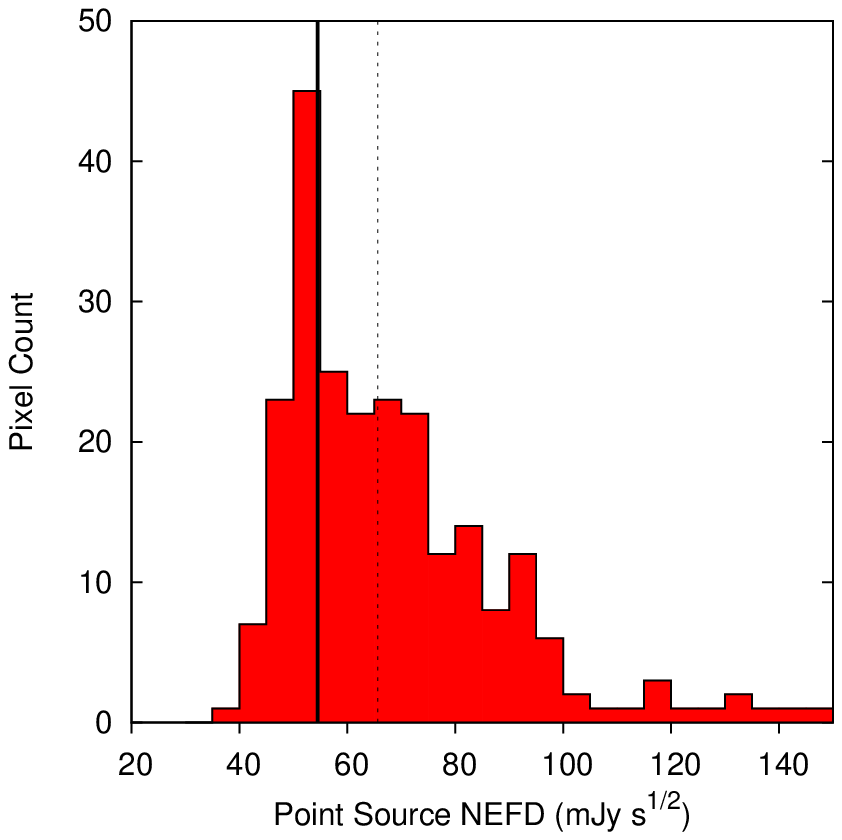} 
\caption{
{\it Top:} Effective point source sensitivities (after low frequencies filtering) of all the useful bolometers of LABOCA
(color scale in Jy\,beam$^{-1}$).  The positions were determined from beam maps on Mars, thus providing a realistic picture of
the optical distortions over the FoV (see also Fig.~\ref{fig:beams}).
{\it Bottom:} Distribution of the effective point source sensitivities on the LABOCA array (5\,mJy\,s$^{1/2}$ binning).
The total mapping speed of the array is as if the 250 good bolometers were all identical at a level of
54.5\,mJy\,beam$^{-1}$\,s$^{1/2}$ (thick black line). The median sensitivity is also shown (dotted line).
}
\label{fig:sensitivity}
\end{figure}

The noise-weighted mean point-source sensitivity of the array
(noise-equivalent flux density, NEFD) determined from on-sky
integrations, is 55\,mJy\,s$^{1/2}$ (sensitivity per channel).  This
value is achieved only by filtering the low frequencies (hence large
scale emission) to reject residual sky-noise.  For extended emission,
without low frequency filtering, there is a degradation of sensitivity
to a mean array sensitivity of 80--100\,mJy\,s$^{1/2}$ depending on
sky stability. However, there are significant variations of the
sensitivity across the array (see Fig.~\ref{fig:sensitivity}, top).

For detection experiments of compact sources with known position,
LABOCA can be centered on the most sensitive part of the array rather
than on the geometric center. This results in an improved point source
sensitivity of $\sim$40\,mJy\,s$^{1/2}$ for compact mapping pattern
like spirals.
\subsection{Mapping speed and time estimate}
The relation between the expected residual rms map noise $\sigma$, the surveyed
area on the sky and the integration time can be expressed as
\begin{equation}
t_{\rm int} = \frac{(X_{\rm scan}+D)~(Y_{\rm scan}+D)}{A_{\rm beam}~N_{\rm bol}} \left(\frac{ f_{\rm samp}~NEFD~~e^{\tau_{\rm los}}}{\sigma}\right)^2~\\
\\
\end{equation}
where $X_{\rm scan}$, $Y_{\rm scan}$ are the dimensions of the area covered by
the scanning pattern, $D$ is the size of the array, $A_{\rm beam}$ is the
area of the LABOCA beam, $N_{\rm bol}$ is the number of working
bolometers, $t_{\rm int}$ is the on source integration time, $NEFD$ is the
average array sensitivity, $f_{\rm samp}$ is the number of grid points
per beam and $\tau_{\rm los}$ the line-of-sight opacity.
This formula\footnote{An online time estimator is available at:\\
\tiny{\url{http://www.apex-telescope.org/bolometer/laboca/obscalc/}}}
does not include sensitivity variations across the array and the increasing
sparseness of data points toward the map edges (the latter effect
depends on the mapping mode).  We have tested this estimate on deep
integrations for point like and extended sources and found a
reasonable agreement in the measured rms noise of the processed maps.
\subsection{Noise behavior in deep integrations}
\begin{figure}[t] 
\centering 
\includegraphics[width=.90\linewidth, bb=0 0 351 357, clip]{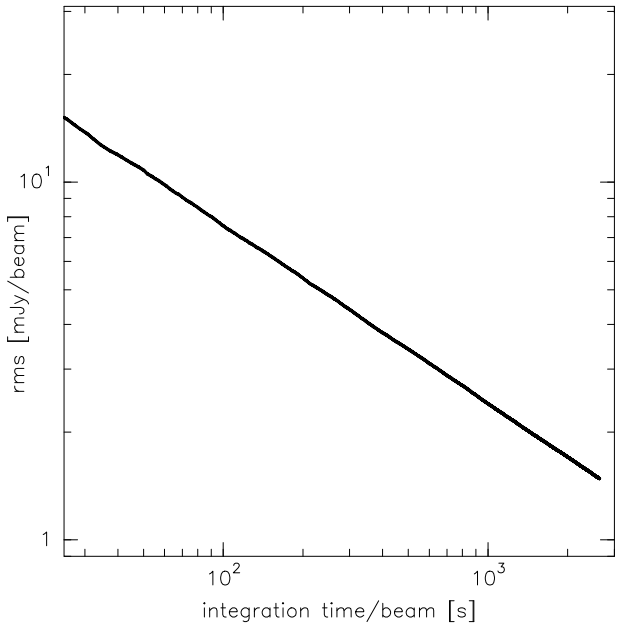} 
\caption
{
Noise behavior for deep integrations has been studied by co-adding about 350 hours of
blank-field data.  The plot shows the effective residual rms noise on sky.
}
\label{fig:noise} 
\end{figure}
In order to study how the noise averages down with increasing
integration time, {about 350 hours of blank-field mapping observations}
have been co-added with randomly inverted signs to remove any sources while
keeping the noise structure.  As shown in~Fig.~\ref{fig:noise}, the
noise integrates down with t$^{-1/2}$, as expected.
%
\section{Data reduction}\label{sec:redu} 
LABOCA data are stored in MB-FITS format (Multi-Beam FITS) by the data
writer embedded in APECS.  A new software package has been
specifically developed to reduce LABOCA data: the Bolometer array data
Analysis software (BoA). It is mostly written in the Python language,
except for the most computing demanding tasks, which are written in
Fortran90.

BoA was first installed and integrated in APECS in early 2006. An extensive
description of its functionalities will be given in a separate paper (Schuller
et al.~in prep.). In this Section, we outline the most important features
for processing LABOCA data.
\subsection{On-line data reduction}
During the observations carried out at APEX, the on-line data
calibrator (as part of APECS) performs a quick data reduction of each
scan to provide the observer with a quick preview of the maps or
spectra being observed.  This is of particular importance for the
basic pointing and focus observing modes, for which the on-line
calibrator computes and sends back to the observer the pointing
offsets or focus corrections to be applied.  For both focus and
pointing scans, only a quick estimate of the correlated noise (see
below) is computed and subtracted from the data.  The focus correction
is derived from a parabolic fit to the peak flux measured by the
reference bolometer as a function of the subreflector position.  For
pointing scans, the signals of all usable channels are combined into a
map of the central $5'\times 5'$ area, in horizontal coordinates.  A
two-dimensional elliptical Gaussian is then fitted to the source in
this map, which gives the pointing offsets, as well as the peak flux
and the dimensions of minor and major axis of the source.
\subsection{Off-line data reduction}
The BoA software can also be used off-line to process any kind of
bolometer data acquired at APEX.  The off-line BoA runs in the
interactive environment of the Python language.  In a typical off-line
data reduction session, several scans can be combined together, for
instance to improve the noise level on deep integrations, or to do
mosaicing of maps covering adjacent areas.  The result of any data
reduction can be stored in a FITS file using standard world coordinate
system (WCS) keywords, which can then be read by other softwares for
further processing (e.g.~source extraction, or overlay with ancillary
data).

The common steps involved in the processing of LABOCA data are the
following: \begin{itemize} \item Flux calibration.  A correct scaling
of the flux involves, at least, two steps: the opacity correction and
the counts-to-Jy conversion.  The zenith opacity is derived from
skydip measurements (see Sect.~\ref{sec:skydips} and
Sect.~\ref{sec:opacity}), and the line of sight opacity also depends
on the elevation.  The counts-to-Jy factor has been determined during
the commissioning of the instrument, but additional correction factors
may be applied, depending on the flux measured on calibrators with
known fluxes (see Sect.~\ref{sec:calibra}).  \item Flagging of bad
channels.  Bolometers not responding or with strong excess noise are
automatically identified from their rms noise being well outside the
main distribution of the rms noise values across the array. They can
be flagged, which means that the signal that they recorded is not used
any further in the processing.  
\begin{figure}[t] 
\centering
\includegraphics[width=.96\linewidth, bb=0 0 340 244,clip]{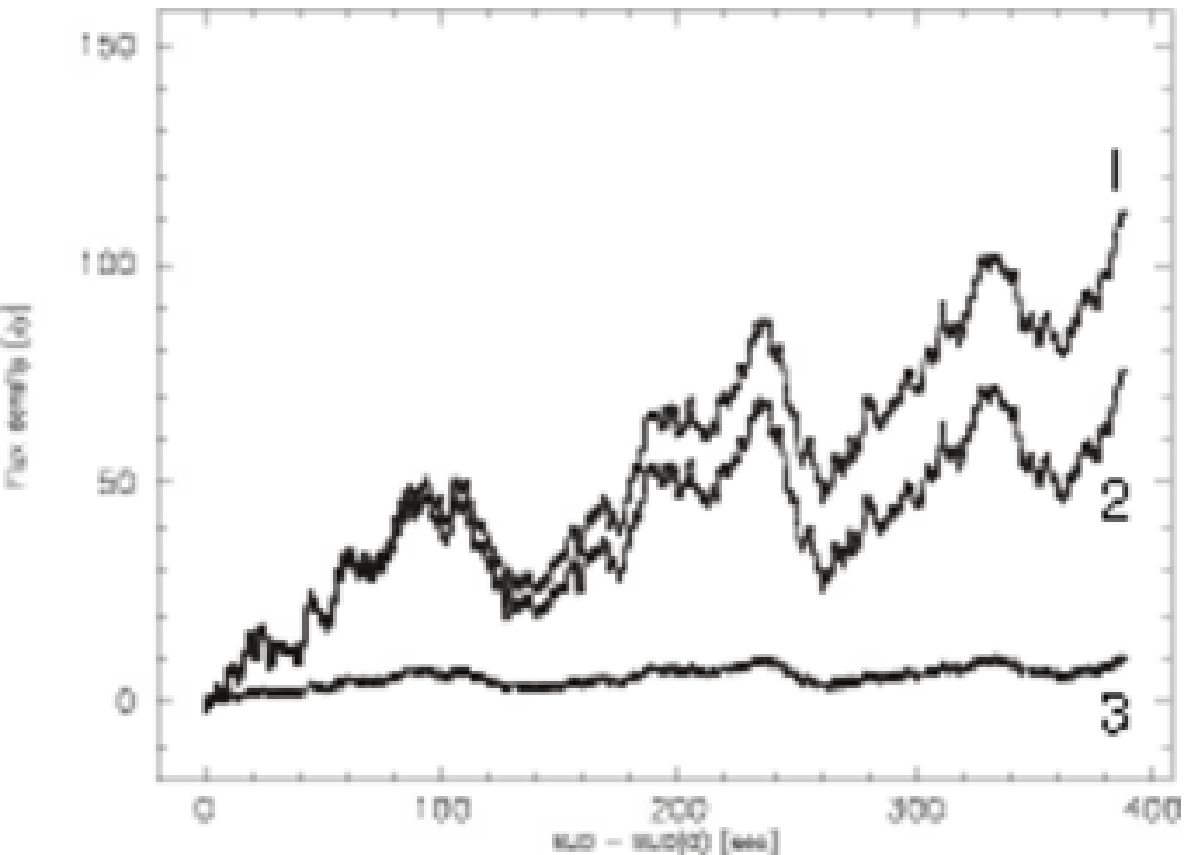} 
\includegraphics[width=.96\linewidth, bb=0 0 339 245, clip]{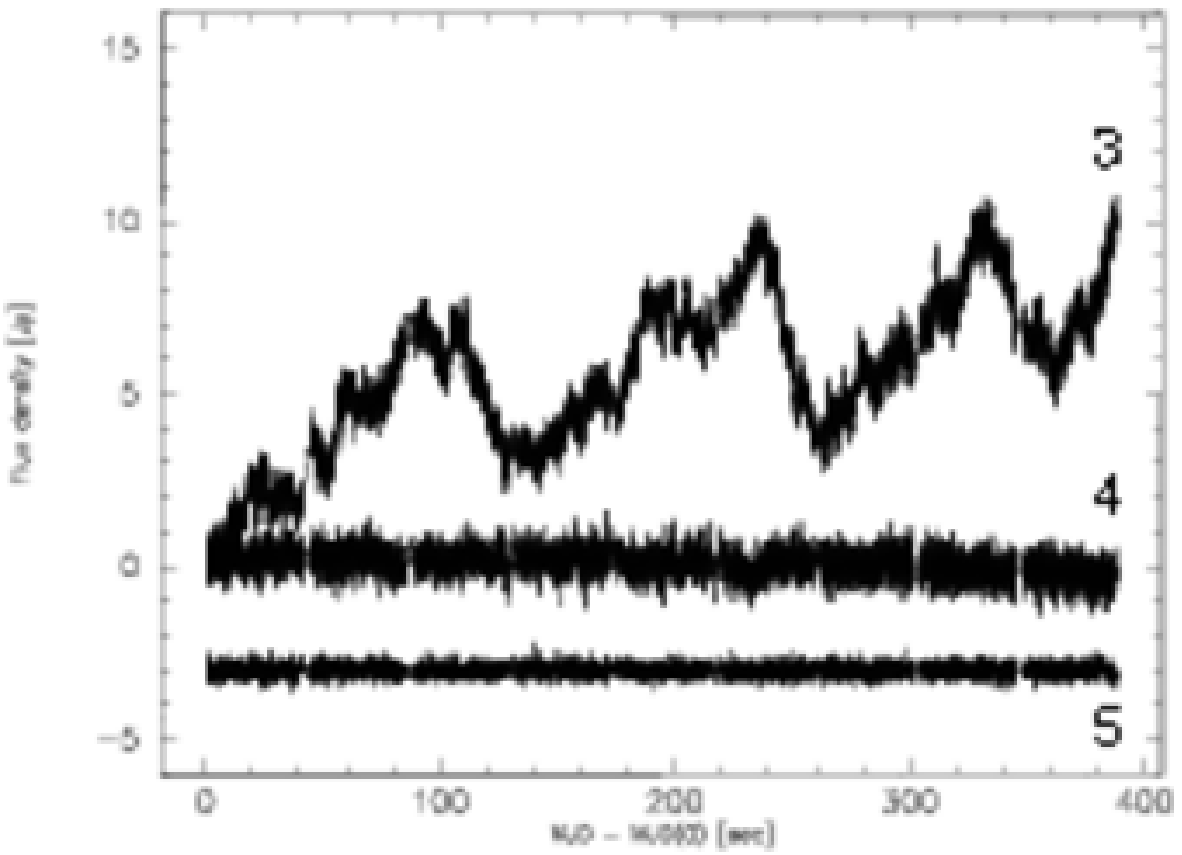} 
\caption
{ 
Time series of a bolometer signal on blank sky at consecutive steps during the data reduction process with BoA.  The signals, labeled from 1 to
5, are shown for:
1) original signal on sky;
2) after correcting for system temperature variations using the blind bolometers ($\sim$200\,\microK during this scan);
3) after median skynoise removal over the full array, one single iteration;
4) after median skynoise removal over the full array, 10 iterations;
5) after correlated noise removal, grouping the channels by amplifier and cable, 5 iterations (shifted to a $-$3 level, for clarity).
}
\label{fig:reduction} 
\end{figure} 
\item Flagging of stationary
points.  The data acquired when the telescope was too slow to produce
a signal inside the useful part of the post-detection frequency band
of LABOCA (e.g.~below 0.1\,Hz, see Sec.~\ref{sec:mapmodes} and
Fig.~\ref{fig:noisespectra}) can be flagged.  Data obtained when the
telescope acceleration is very high may show some excess noise, and
can also be flagged.  
\item Correlated noise removal.  This can be
done using a Principal Components Analysis (PCA), or a median noise
removal method. In the latter case, the median value of all
(normalized) signals is computed at each timestamp, and subtracted
from the signal of each channel (with appropriate relative gains).
This can be performed using all beams at once or better on groups of
selected beams.  In fact, some groups of channels sharing the same
electronics subsystem (e.g.~amplifier box, flat cable) can show strong correlation
and removing the median signal on those groups of channels greatly improves
their signal-to-noise ratio.  However, it should be noted that an astronomical source
with extended uniform emission on scales of several arcmin or larger would mimic the
correlated noise behavior of groups of channels. Therefore,
subtracting the median noise results in filtering out some fraction of
the extended emission, depending on the size and morphology of the source.  
\item Despiking. Data points that deviate by more than a
given factor times the standard deviation in each channel can be flagged.  
\item Data weighting and map making.  To build a map in
horizontal or equatorial coordinates, the data of all usable channels
are projected onto a regular grid and co-added, using a weighted
average, with weights computed as 1/rms$_{\rm c}^2$, where rms$_{\rm c}$
refers to the rms noise of an individual channel. The channel rms
noise can either be the standard deviation of the data over the full
time line, or it can be computed on a sliding window containing a
given number of integrations.  \end{itemize} A visual example of
correlated noise removal with BoA is given
in~Fig.~\ref{fig:reduction}.

Additional (optional) steps that can improve the final reduction
include: removing slow variations from the signal, by subtracting a
polynomial baseline or by filtering out low frequencies in the Fourier
domain; smoothing of the map with a Gaussian kernel of a given size.

The map resulting from a full reduction can be used as a model of the
astronomical source to mask the data, in order to repeat some
computations without using data points corresponding to the source in
the map.  This iterative scheme helps in the difficult task to recover
the extended emission and reduces the generation of artifacts around
strong sources.  In the presence of bright sources in the map, a
typical data reduction should first perform the reduction steps as
listed above with conservative parameters (for example, using high
enough thresholds in the despiking to avoid flagging out the strong
sources), and should then read again the raw data, use the resulting
map to temporarily flag out datapoints corresponding to bright
sources, and repeat the full process with less conservative
parameters.

In addition to the data processing itself, BoA allows visualization of
the data in different ways (signal vs. time, correlation between channels,
power spectra or datagrams), as well as the telescope pattern, speed
and acceleration.  Finally, BoA also includes a simulation module,
which can be used to investigate the mapping coverage for on-the-fly
maps, spirals and raster of spirals, given the bolometer array
parameters.
%
%
\section{Science with LABOCA}\label{sec:science}
Because of its spectral passband, centered at the wavelength of
870\,\micron (see Fig.~\ref{fig:passband}), LABOCA is particularly
sensitive to thermal emission from cold objects in the Universe which
is of great interest for a number of astrophysical research fields.
\subsection{Planet Formation}
The study of Kuiper Belt Objects in the Solar System as well as
observations of debris disks of cold dust around nearby main sequence
stars can give vital clues to the formation of our own planetary
system and planets in general.  With its angular resolution of~19\arcsec
(see Fig.~\ref{fig:beamshape}) LABOCA can resolve the debris disks of
some nearby stars.
\subsection{Star Formation in the Milky Way}
\begin{figure}[t]
\centering
\includegraphics[width=.99\linewidth,bb=20 20 432 786, clip]{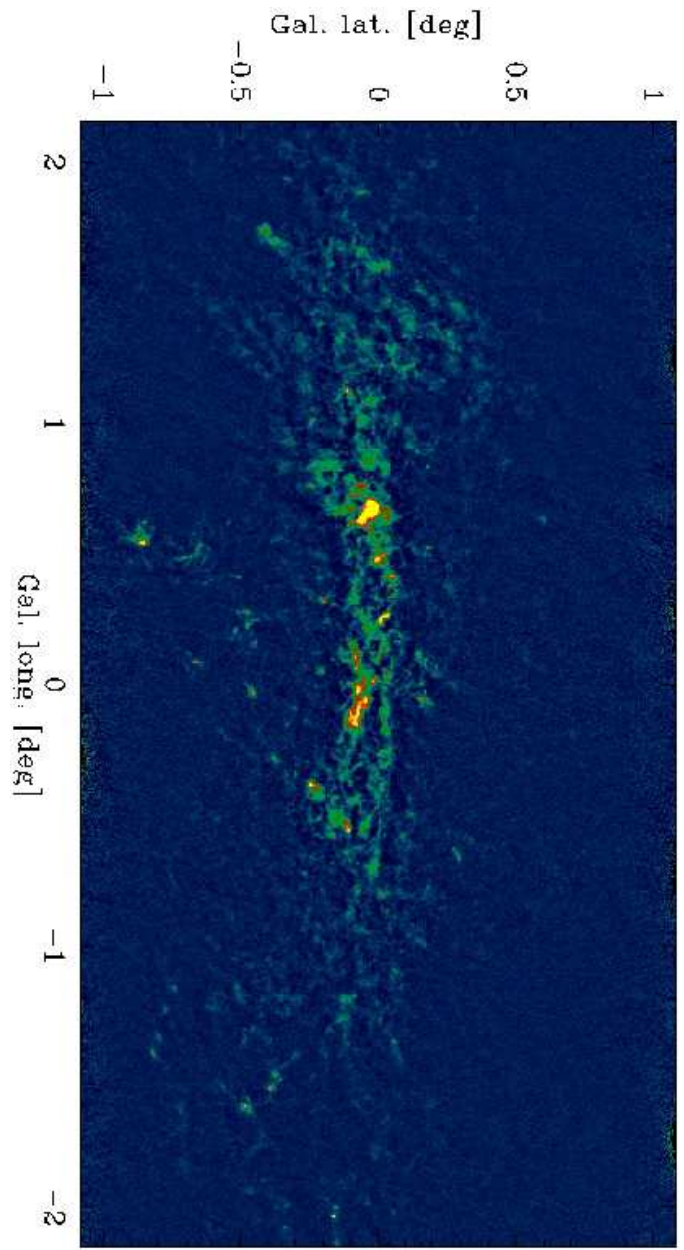} 
\caption
{
A map of the galactic center, extracted from the project ATLASGAL (F. Schuller et al., in prep.).
The map is 2$\times$4 degrees large, the residual noise is 50\,mJy\,beam $^{-1}$ and it required only 6~hours of
observing time.  Data reduced with the BoA package.
}
\label{fig:atlasgal}
\end{figure}
The outstanding power of LABOCA in mapping large areas of sky with
high sensitivity (see Fig.~\ref{fig:atlasgal}) makes it possible,
for the first time, to perform unbiased surveys of the distribution
of the cold dust in the Milky Way.
As the dust emission at 870\,\micron  is typically optically thin, it is
a direct tracer of the gas column density and gas mass.  Large scale
surveys in the Milky Way will reveal the distribution and gas
properties of a large number of pre-star cluster clumps and
pre-stellar cores in different environments and evolutionary
states. Equally importantly, they provide information on the structure
of the interstellar medium on large scales at high spatial resolution,
little explored so far.  Such surveys are vital to improve our
understanding of the processes that govern star formation as well as
the relation between the clump mass spectrum and the stellar initial
mass function.  Large unbiased surveys are also critical for finding
precursors of high-mass stars, which are undetectable at other
wavelengths due to the high obscuration of the massive cores they are
embedded in. LABOCA will help to obtain a detailed understanding of
their evolution.  In addition, deep surveys of nearby, star-forming
clouds, will allow study of the pre-stellar mass function down to the
brown dwarf regime.
\subsection{Cold gas in Galaxies}
\begin{figure}[t] 
\centering
\includegraphics[width=.99\linewidth, bb=0 0 416 307, clip]{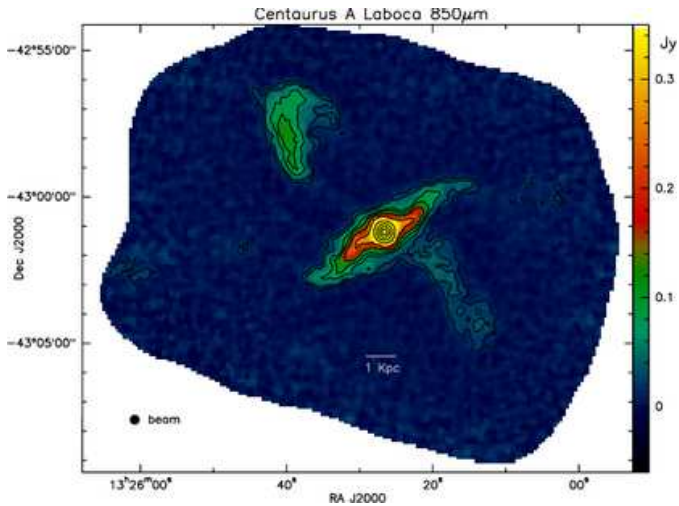} 
\caption
{
Map of the nearby galaxy Centaurus A \citep[\object{NGC 5128}, from ][]{2008A&A...490...77W}.
The central dust disk has a pronounced S-shape.  Our LABOCA image shows, for the first time at
submillimeter wavelengths, the synchrotron emission from the radio jets and
the inner radio lobes.  Its distribution follows closely the known
radio emission at cm wavelengths.  The color scale is in units
of Jy\,beam$^{-1}$. The residual rms noise is 4\,mJy\,beam$^{-1}$ and
the total integration time is 5~hours.
}
\label{fig:ngc5128}
\end{figure}

The only reliable way to trace the bulk of dust in galaxies is through
imaging at submillimeter wavelengths.  It is becoming clear that most
of the dust mass in spiral galaxies lies in cold, low
surface-brightness disks, often extending far from the galactic
nucleus, as in the case of the starburst galaxy \object{NGC 253} or of \object{NGC 5128}
\citep[\object{Cen A}, see Fig.~\ref{fig:ngc5128} and][]{2008A&A...490...77W}.
Understanding this component is critically important as
it dominates the total gas mass in galaxies and studies of the Schmidt
law based on \ion{H}{i} observations only can heavily underestimate
the gas surface density in the outer parts.

In addition to studying individual nearby galaxies, LABOCA will be
vital for determining the low-$z$ benchmarks, such as the local
luminosity and dust mass functions, which are required to interpret
information from deep cosmological surveys.
\subsection{Galaxy formation at high redshift}
\begin{figure}[t]
\centering
\includegraphics[width=.99\linewidth, bb=28 192 567 610, clip]{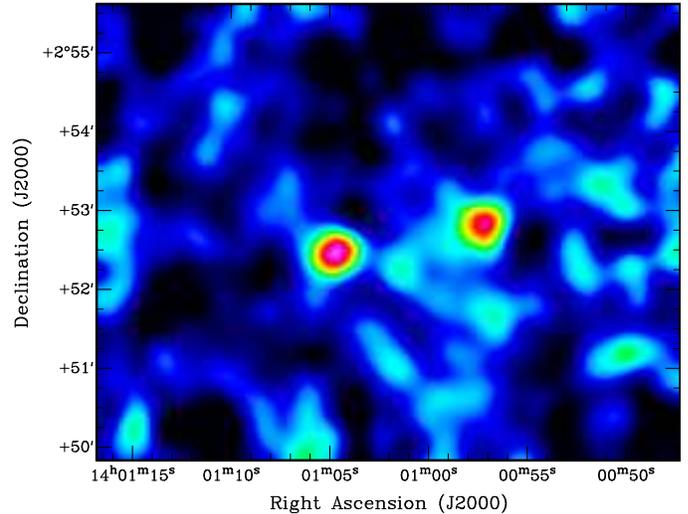}
\caption
{
LABOCA images of \object{SMM J14011+0252} (left) and \object{SMM J14009+0252} (right) smoothed to $25''$ resolution.
SMM\,J14011 is at a redshift of $z=2.56$ while SMM\,J14009 has no reliable redshift determination.
The noise level of the map is about 2.5\,mJy\,beam$^{-1}$.
}
\label{fig:smm14009}
\end{figure}

Due to the negative-K correction, submillimeter observations offer
equal sensitivity to dusty star-forming galaxies over a redshift range
from $z\sim$1--10 and therefore provide information on the star
formation history at epochs from about half to only 5\% of the present
age of the universe. Recent studies have shown that the volume density
of luminous {\it submillimeter galaxies} (SMGs) increases over a
thousand-fold out to $z\sim$2 \citep{2005ApJ...622..772C}, and thus,
in contrast to the local Universe, luminous obscured galaxies at high
redshift could dominate the total bolometric emission from all
galaxies at early epochs. The mass-tracing property of submillimeter
dust emission (see Sect.~\ref{sec:intro}) allows us to make a direct estimate
of the star formation rates (SFRs) of these objects. The generally high
observed SFRs suggest that approximately half of all the stars that
have formed by the present day may have formed in highly obscured
systems which remain undetected at optical or near-infrared
wavelengths. One example for such system is the submillimeter galaxy
\object{SMM J14009+0252} (see Fig.~\ref{fig:smm14009}), which is strong in the
submillimeter range, has a 1.4\,GHz radio counterpart, but no obvious
counterpart in deep K-band images \citep{2000MNRAS.315..209I}.
Clearly it is critical to include these highly obscured sources in
models of galaxy formation to obtain a complete understanding of the
evolution of galaxies.
With its fast mapping capabilities LABOCA allows us to map fields of a
half square degree, a typical size of a deep cosmological field
surveyed at other wavelengths, down to the confusion limit in a
reasonable amount of observing time. This will also greatly improve
the statistics of high redshift galaxies detected at submillimeter
wavelengths. See \citet{Beelen2008} for the first deep LABOCA
cosmology imaging.
%
%
\section{Future plans}\label{sec:future}
In collaboration with the Institute for Photonics Technology (IPHT, Jena, Germany)
we are working on a new bolometer array, LABOCA-2, using
superconducting technology.  The new array will use superconducting
thermistors (transition edge sensor, TES), planar dipole absorbers,
SQUID (Superconducting Quantum Interference Device) multiplexing and
amplification.  A system already using the same technology, called
Submillimeter APEX Bolometer Camera (SABOCA) is going to be
commissioned as facility instrument on the APEX telescope for
operation in the 350\,\micron  atmospheric window.  Additionally, given
the low sensitivity of superconducting bolometers to microphonics, it
will be possible to move from the wet cryostat to a pulse-tube cooler,
as already successfully tested at MPIfR in Bonn, improving
considerably the conditions of operability of the system.
%
%
\begin{acknowledgements}
The authors would like to thank the staff at the APEX telescope for their
support during installation and commissioning of the instrument.  Special
thanks to Lars-{\AA}ke Nyman (ESO, APEX station manager at the time of the
commissioning) for his invaluable help and for his belief in the project
even in the most difficult moments.
E.K. enjoyed the hospitality and help provided by the staff and many student members
of the Microfabrication Facility of UC Berkeley during the manufacture of the LABOCA wafer.
\end{acknowledgements} 
%
%
%
%
\appendix
\section{Secondary calibrators}\label{sec:appendix}
\subsection{The absolute flux scale and secondary calibrators} 
The absolute flux density calibration of (sub)millimeter photometry is
usually established via observations of planets with
well characterized temperature models and angular sizes smaller than
the telescope beam ($19''$ for LABOCA on APEX, see \ref{sec:beam}). Thus, the primary
calibrators for LABOCA are Mars, Uranus, and Neptune. A planet's flux density is,
to first order, determined by its angular size and its temperature at
the time of observation, which in turn is determined by its insolation.
The situation is complicated, a.o., by diurnal temperature variations and
the physics and composition of the planetary atmospheres
\citep[see, e.g.,][]{Ulich1981, Hildebrand1985, Griffin1986, Orton1986}.

In practice good calibration of science data requires frequent comparisons to calibration sources,
preferably near the science target, to eliminate systematic calibration effects
(mostly elevation dependent).
For this reason, the small number of primary calibrators
(only three for LABOCA), cannot serve all scientific calibration needs.
It is therefore highly desirable to have a list of additional
compact objects that serve as {\it secondary} calibrators provided their
fluxes have been determined with particular care via comparisons to
the primary calibration sources. A variety of secondary calibrators, with flux densities determined at various wavelengths
350\,\micron\,$ < \lambda < 2\,$mm, using the 15\,m James Clerk Maxwell Telescope (JCMT),
has been published by \citet{Sandell1994}.
\subsection{LABOCA observations of secondary calibrators}
\begin{figure*}[t]
\centering
\includegraphics[width=.99\linewidth, bb=20 20 599 684, clip]{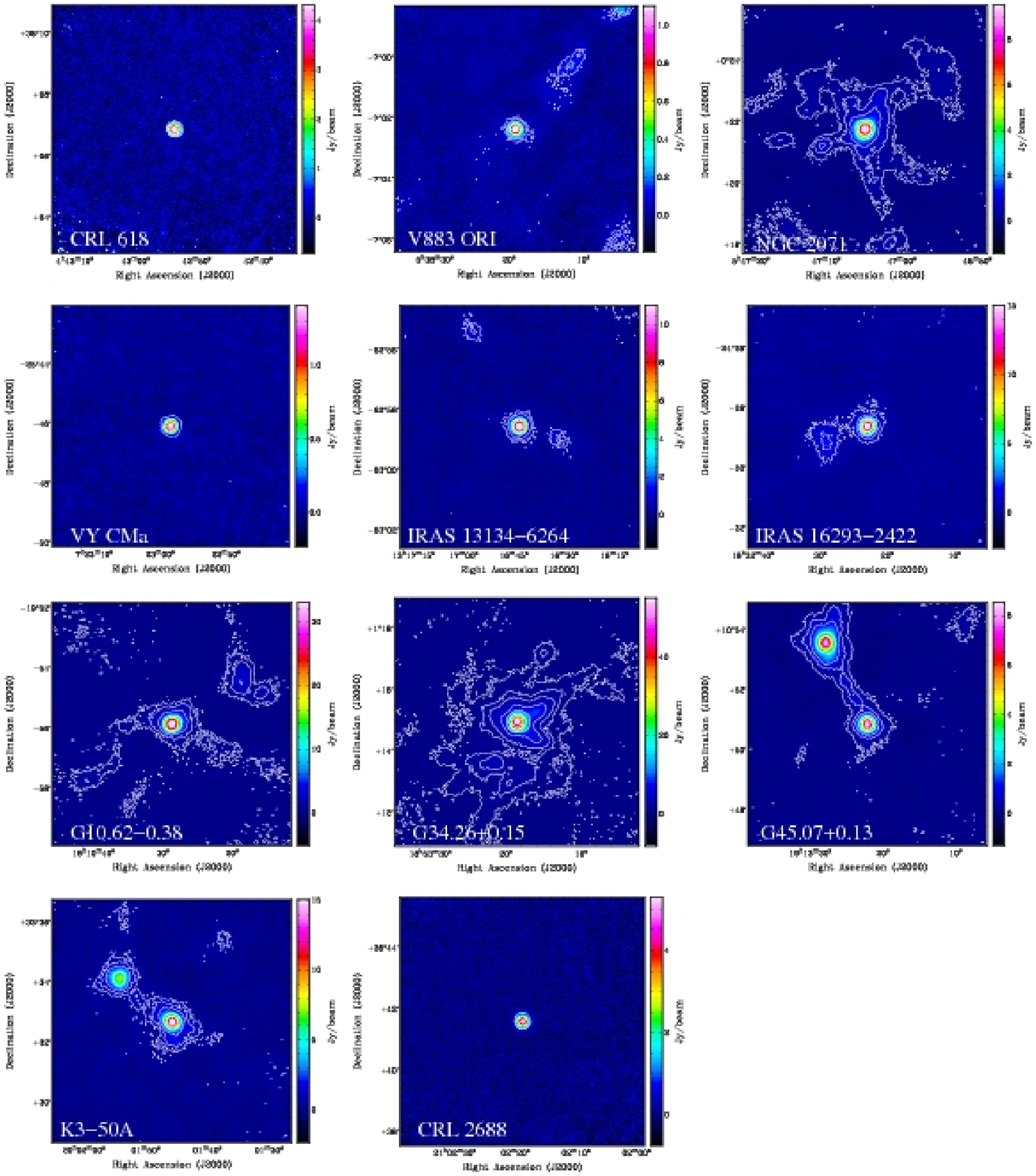}
\caption
{
Maps of the secondary calibrators imaged with LABOCA (ordered in right ascension).
Contours are shown at:
0.75, 1.5, 3.0 Jy beam$^{-1}$ for \object{CRL 618};
0.08, 0.15, 0.5, 1.0 Jy beam$^{-1}$ for \object{V883 ORI};
0.1, 0.4, 1.0, 3.0, 5.0, 8.5 Jy beam$^{-1}$ for \object{NGC 2071};
0.1, 0.4, 1.0, 1.4  Jy beam$^{-1}$ for \object{VY CMa};
0.5, 1.0, 3.0, 5.0, 11.0 Jy beam$^{-1}$ for \object{IRAS 13134-6264};
0.6, 1.5, 3.0, 5.0, 10.0, 15.0 Jy beam$^{-1}$ for \object{IRAS 16293-2422};
0.3, 1.0 3.0 10.0, 30.0 Jy beam$^{-1}$ for \object{G10.62-0.38};
0.2, 1.0, 2.0, 5.0, 10.0, 20.0, 40.0, 55.0 Jy beam$^{-1}$ for \object{G34.26+0.15};
0.13, 0.5, 1.0 2.5, 5.0, 8.0 Jy beam$^{-1}$ for \object{G45.12+0.13};
0.5, 1.0, 2.0, 5.0, 10.0, 15.0 Jy beam$^{-1}$ for \object{K3-50A};
1.0 3.0, 5.0 Jy beam$^{-1}$ for \object{CRL 2688}.
}
\label{fig:calibrators}
\end{figure*}

Since its commissioning, in May 2007, we have used LABOCA at APEX to
observe a number of sources which serve as secondary calibrators.
The sample, which is partially based on \citet{Sandell1994}, consists
of ultracompact H{\scriptsize II} regions, protostellar objects, and
AGB stars. LABOCA observations of secondary calibrators
are typically done in raster-spiral mode leading to a fully sampled map
for each scan. In Fig.~\ref{fig:calibrators} we present images centered
on the LABOCA secondary calibrators, which have been calculated by averaging
several raster-spiral scans for each source.
In Table~\ref{tab:calibrators} we summarize their properties.

For each scan, we fitted elliptical Gaussian profiles to the observed flux density distribution.
The flux values quoted in the table are averages, while the uncertainties represent the rms scatter of the individually fitted peaks.
The fitting errors are generally much smaller than the intrinsic calibration scatter.
The quoted uncertainties, therefore, represent the typical LABOCA calibration accuracy including systematic effects
such as uncertainties in the focus and errors propagated from the opacity determination.

The typical uncertainty of the secondary calibrator fluxes is about 7\%,
comparable to the 8\% scatter on the primary calibrators
(namely Mars, Uranus and Neptune, $\sim$700 observations)
derived in the same way and using the flux predictions from the ASTRO model (see Sect.~\ref{sec:calibra}).
We are thus confident that the calibrator fluxes listed in Table~\ref{tab:calibrators} are accurate to $\sim$10\%.

In addition to the averaged Gaussian peak flux densities (brightness) we also
list, for each calibrator source, the peak flux density measured on the co-added maps
(which differs from the Gaussian peak value for extended, structured sources),
the flux integrated over a Gaussian fitted to the images shown in Fig.~\ref{fig:calibrators},
the apparent major and minor axes and the mean position angles.
\begin{table*}[t]\label{tab:calibrators}
\caption{Properties of the secondary calibrators defined with LABOCA}
\begin{center}
\begin{tabular}{lrrrrrrrrr}
Source & $\alpha({\rm J}2000)$ & $\delta({\rm J2000})$ & \multicolumn{1}{c}{S$^{(a)}_{\rm G}$} & \multicolumn{1}{c}{Nr.} 
& \multicolumn{1}{c}{S$^{(b)}_{\rm peak}$}&$\smallint S{\rm d}\Omega^{(c)}_{\rm G}$&$\theta_{\rm maj}$& $\theta_{\rm min}$ & P.A. \\

&&&\multicolumn{1}{r}{(Jy~beam$^{-1}$)}& &\multicolumn{1}{r}{(Jy~beam$^{-1}$)}& \multicolumn{1}{c}{(Jy)} &($''$) & ($''$) & ($^{\circ}$ E of N)\\
\hline 
CRL 618           & $04^{\rm h}42^{\rm m}54\rlap{.}{^{\rm s}}00$ & $+36^{\circ}06'53\rlap{.}''7$ &  $4.9  \pm 0.2 $ &   6 &  $4.7(2)   $& 5.5(4)  &  19.7(8) & 19.0(8) & --     \\
V883 ORI          & $05^{\rm h}38^{\rm m}18\rlap{.}{^{\rm s}}10$ & $-07^{\circ}02'27\rlap{.}''0$ &  $1.36 \pm 0.11$ &  41 &  $1.35(3)$  & 1.93(6) &  22.3(5) & 21.3(5) &  51(20)\\
NGC 2071          & $05^{\rm h}47^{\rm m}04\rlap{.}{^{\rm s}}85$ & $+00^{\circ}21'47\rlap{.}''1$ &  $9.1  \pm 0.6 $ & 102 &  $10.0(1)$  & 26.6(3) &  30.3(3) & 28.1(3) & 157(4) \\
VY CMa            & $07^{\rm h}22^{\rm m}58\rlap{.}{^{\rm s}}33$ & $-25^{\circ}46'03\rlap{.}''2$ &  $1.54 \pm 0.17$ &  34 &  $1.52(2)$  & 1.90(5) &  21.1(3) & 21.0(3) & --     \\
IRAS 13134$-$6264 & $13^{\rm h}16^{\rm m}43\rlap{.}{^{\rm s}}15$ & $-62^{\circ}58'31\rlap{.}''6$ &  $12.9 \pm 0.9 $ & 134 &  $12.9(1)$& 21.1(2)   &  23.5(2) & 23.3(2) &  --\\
IRAS 16293$-$2422 & $16^{\rm h}32^{\rm m}22\rlap{.}{^{\rm s}}90$ & $-24^{\circ}28'35\rlap{.}''6$ &  $16.4 \pm 1.2 $ & 142 &  $17.4(2)$& 28.5(3)   &  26.9(2) & 24.6(2) & 148(3) \\
G 10.62$-$0.38    & $18^{\rm h}10^{\rm m}28\rlap{.}{^{\rm s}}66$ & $-19^{\circ}55'49\rlap{.}''7$ &  $33.4 \pm 2.0 $ & 408 &  $35.4(2)$& 72.0(6)   &  29.2(2) & 24.5(1) & 100(1) \\
G 34.26$+$0.15    & $18^{\rm h}53^{\rm m}18\rlap{.}{^{\rm s}}50$ & $+01^{\circ}14'58\rlap{.}''6$ &  $55.5 \pm 3.5 $ & 164 &  $60.1(1)$  & 128.5(6)&  27.7(1) & 26.2(1) & 103(3) \\
G 45.07$+$0.13    & $19^{\rm h}13^{\rm m}22\rlap{.}{^{\rm s}}07$ & $+10^{\circ}50'53\rlap{.}''4$ &  $8.2  \pm 0.6 $ &  73 &  $8.50(5)$  & 15.9(2) &  26.2(2) & 24.6(2) &  67(4) \\
K3-50A            & $20^{\rm h}01^{\rm m}45\rlap{.}{^{\rm s}}69$ & $+33^{\circ}32'43\rlap{.}''5$ &  $15.0 \pm 1.2 $ &  23 &  $16.6(2)$  & 37.7(9) &  31.1(4) & 26.9(3) &  49(4) \\
CRL 2688          & $21^{\rm h}02^{\rm m}18\rlap{.}{^{\rm s}}80$ & $+36^{\circ}41'37\rlap{.}''7$ &  $5.4  \pm 0.4 $ &  20 &  $5.3(2)$   & 6.6(1)  &  23.2(2) & 22.5(2) &  53(14)\\
\hline\\
\multicolumn{10}{l}{$^{(a)}$Mean peak flux derived from a Gaussian fit to all individual scans, errors are the standard deviation of these values.}\\
\multicolumn{10}{l}{$^{(b)}$Peak flux on the co-added map (given in Fig.~\ref{fig:calibrators}), errors correspond to the map rms.}\\
\multicolumn{10}{l}{$^{(c)}$Integrated flux of a Gaussian fit on the co-added map, errors are the formal fitting errors.}\\
\end{tabular}
\end{center}
Columns are (left to right) source name, right ascension and declination (equinox J2000), Gaussian peak flux,
number of observations, map peak flux, integrated Gaussian flux, major and the minor axis and mean position angle (east of north).
The error for the Gaussian peak flux is the standard deviation of all measurements
and gives the typical LABOCA calibration accuracy including systematic effects such as uncertainties
in the focus and errors propagated from the opacity determination.
The numbers in parentheses are formal fitting uncertainties in the last quoted digit.
The overall uncertainties are estimated to be $\sim$10\%,
including the uncertainties in the observations of the primary calibrators (Mars, Uranus, Neptune).
\end{table*}
\subsection{Discussion}
Most of the secondary calibrators defined with LABOCA are also included in the study of \citet{Sandell1994},
where photometric flux densities in the JCMT beam are quoted.
At the wavelength $\lambda = \,850\,\micron$,
the closest to the LABOCA's center wavelength among the ones used by Sandell,
the reported beam is $17\rlap{.}''5$, comparable to LABOCA's $19''$ (see Sect.~\ref{sec:passband}).

Comparison of the peak flux densities measured with LABOCA and the ones from \citet{Sandell1994} shows remarkable agreement.
Our peak flux densities on the maps
(which is the appropriate value for comparison to the photometric flux, in particular, for non-Gaussian, extended sources)
agree with photometric values given by Sandell to within 5\% on average (excluding \object{VY CMa}).
\object{VY CMa} is, in fact, known to be variable \citep{knapp93} and our 2007--2008 measurements
show flux densities about 40\% lower than the values observed in 1989--1991 by Sandell.

The secondary calibrators all have some line emission (e.g. CO(3-2)), which might be a significant fraction of the total for some of them.
The line contamination in the fluxes measured by LABOCA, however, has no impact on the calibration of LABOCA science data.
%
%
\bibliographystyle{aa} \bibliography{1454light} 
\end{document}